\documentclass[a4paper,fleqn,usenatbib]{mnras}

\usepackage{newtxtext,newtxmath}
\usepackage[T1]{fontenc}
\usepackage{ae,aecompl}

\usepackage{graphicx}	% Including figure files
\usepackage{amsmath}	% Advanced maths commands
\usepackage{rotating}

\newcommand{\Eexc}{$E_{\rm exc}$}
\newcommand*\rfrac[2]{{}^{#1}\!/_{#2}}
\newcommand{\Teff}{T_{\rm eff}}
\newcommand{\logg}{\rm log~ g}
\newcommand{\kms}{km\,s$^{-1}$}
\newcommand{\Ebv}{E_{B-V}}
\newcommand{\eps}[1]{\log\varepsilon_{\rm #1}}

\def\ione{\,{\sc i}}
\def\ii{\,{\sc ii}}

\title[Non-LTE abundance analysis of the Sextans galaxy]{The formation of the Milky Way halo and its dwarf satellites: \\ A NLTE-1D abundance analysis. V. The Sextans galaxy\thanks{Based on UVES/VLT observations collected at the European South Observatory and data collected at Subaru Telescope and obtained from the SMOKA, which is operated by the Astronomy Data Center, National Astronomical Observatory of Japan.}}

\author[L. Mashonkina et al.]{
L.~Mashonkina,$^{1}$\thanks{E-mail: lima@inasan.ru} 
Yu.~V.~Pakhomov,$^{1}$ 
T.~Sitnova,$^{1}$ 
P.~Jablonka,$^{2,3}$ 
\newauthor S.~A. Yakovleva,$^{4}$ 
and A.~K. Belyaev,$^{4}$  \\
$^{1}$Institute of Astronomy of the Russian Academy of Sciences, Pyatnitskaya st. 48, 119017, Moscow, Russia \\
$^{2}$ Laboratoire d' Astrophysique, Ecole Polytechnique F\'ed\'erale de Lausanne (EPFL), Observatoire de Sauverny, CH-1290 Versoix, Switzerland \\
$^{3}$ GEPI, Observatoire de Paris, CNRS, Universit\'e Paris Diderot, F-92125 Meudon Cedex, France \\
$^{4}$ Department of Theoretical Physics and Astronomy, Herzen University, St. Petersburg 191186, Russia 
}

\date{Accepted XXX. Received YYY; in original form ZZZ}

\pubyear{2021}

\begin{document}
\label{firstpage}
\pagerange{\pageref{firstpage}--\pageref{lastpage}}
\maketitle

\begin{abstract}
We present a homogeneous set of accurate atmospheric parameters for a sample of eleven very metal-poor ($-3.32 \le$ [Fe/H] $\le -2.61$) stars in the Sextans dwarf spheroidal galaxy (dSph) and the non-local thermodynamic equilibrium (NLTE) abundances of, at least, seven chemical elements based on high-resolution UVES/VLT and HDS/Subaru spectra. For each star, its effective temperature and surface gravity were derived from the spectral energy distribution and the known distance, and the Fe abundance was obtained as the average from lines of Fe\ione\ and Fe\ii. Based on abundances of Mg, Ca, and Ti, we show that all the investigated stars reveal $\alpha$-enhancements of 0.4~dex to 0.2~dex, and there is a hint of a decline in $\alpha$/Fe for [Fe/H] $> -2.8$. 
The Sextans stars are deficient in Ba.
The new result is an extremely tight relation between Ba and Mg, suggesting their common origin in massive stars and Ba synthesis
%. This means that Ba was produced 
in the r-process events on the timescales of standard supernovae. The exception is a C-enhanced star S~15-19 which is strongly enhanced in Ba. This star is unlikely to be a CEMP-s star because of low abundances of Sr and Y ([Y/Fe] $< -1$) that are produced in the s-process as efficiently as Ba and a non-detection of variation in the radial velocity.
No distinctions from the Milky Way halo and the Sculptor and Ursa Minor dSphs were found in a history of early enrichment in Na and Ni, namely, the Sextans stars are deficient in Na and have close-to-solar Ni/Fe. 
\end{abstract}


%%%%%%%%%%%%%%%%% BODY OF PAPER %%%%%%%%%%%%%%%%%%

\section{Introduction}

The nature and history of the Milky Way (MW) are closely connected to its dwarf satellites via common origin, tidal interactions, and mergers. A merger of the Sagittarius dwarf galaxy with the MW takes place before our eyes \citep{1994Natur.370..194I}. Stellar streams uncovered near the Sun are indicative of some past mergers \citep{1999Natur.402...53H}.
% which can be.
Our understanding of the formation of the MW is dramatically changing in recent years thanks to successful space mission Gaia \citep{2016A&A...595A...1G,2018AA...616A...1G} and large stellar spectroscopic surveys, such as APOGEE \citep{2017AJ....154...94M}, RAVE \citep{2017AJ....153...75K}, LAMOST \citep{2012RAA....12..735D}, and GALAH \citep{2015MNRAS.449.2604D}. For example, 
 the inner halo is found to be dominated by debris from a galaxy named Gaia-Enceladus (or Gaia Sausage), which was as massive as the Small Magellanic Cloud and accreted by the MW around 10 Gyr ago \citep{2018MNRAS.478..611B,2018Natur.563...85H}. \citet{2019MNRAS.488.1235M} identify the Sequoia merger event. \citet{2019A&A...631L...9K} report a separate structure named Thamnos that can be the debris from the accreted small galaxy. 
%The substructures identified by \citet{2018ApJ...860L..11K,2019A&A...631L...9K} are associated to the debris of dwarf galaxies named Sequoia and Thamnos. 
In the view of an impotant role of dwarf galaxies as building blocks of the MW, stellar populations of the surviving dwarfs, in particular, an early history of their star formation, chemical enrichment, and mixing processes should be thoroughly investigated. 

%\citep[see][for a review]{2020ARA&A..58..205H}. 
%Abolfathi,Astrophys. J. Suppl., 235, 42 (2018)
%Studying the Milky Way’s dwarf galaxy satellites offers the opportunity to learn about galaxy formation on small scales. Their  i Chemical abundances of their old stellar populations reflect an n 

For this study, we selected the Sextans dwarf spheroidal galaxy (dSph). It is composed of predominantly old stellar population. According to \citet{2011MNRAS.411.1013B}, the majority of the Sextans stars have [Fe/H] values between $-3.2$ and $-1.4$, with the average [Fe/H] = $-1.9$, which supports the first estimate of the mean metallicity, [Fe/H] = $-2.05$, by \citet{1993ApJ...418..208S}.

 The first high-resolution ($R = \lambda/\Delta\lambda \simeq$ 34\,000) spectral observations were performed by \citet{2001ApJ...548..592S}, for the five stars. That study provides some evidence for a trend of decreasing [$\alpha$/Fe] abundance with increasing metallicity. The [Mg/Fe] ratios were found to vary from a subsolar value of $-0.46$ in the most metal-rich star ([Fe/H] = $-1.45$) up to [Mg/Fe] = 0.41 in the most metal-poor star ([Fe/H] = $-2.85$, S~49). Taking into account the three measured $\alpha$-process elements Mg, Ca, and Ti, \citet{2001ApJ...548..592S} obtain [$\alpha$/Fe] = 0.02 for the Sextans dSph, on average. This is lower than the corresponding value for the MW halo field stellar samples; see, for example, \citet{1995AJ....109.2757M,2004AJ....128.1177V,Cohen2013}, and \citet{2019ARep...63..726M}. The latter paper reports on [$\alpha$/Fe] $\simeq$ 0.3 for each of the Mg, Ca, and Ti elements in the MW halo stars with [Fe/H] $< -1$. 
%, [$\alpha$/Fe] = 0.28,
% reveals an underabundance of the , while  has the highest enhancement of Mg, with .
%[Mg/Fe] = 0.27,0.23, 0.41, -0.46, -0.07; [Ca/Fe] = 0.11,0.08, 0.08, -0.12, 0.34.

A behavior of [$\alpha$/Fe] with metallicity in a galaxy traces its star-formation history. Iron is produced not only in type~II supernovae (SNe~II), together with the $\alpha$-elements, but also in type~Ia supernovae (SNe~Ia) that have longer time scale than SNe~II \citep{1979ApJ...229.1046T}. Therefore, the stars with enhanced $\alpha$-elements trace the epoch of dominance of massive stars in nucleosynthesis, while the stars with decreasing [$\alpha$/Fe] the epoch, when the contribution of SNe~Ia counter balance sufficiently the core-collapse SNe nucleosynthesis products. In case of good enough stellar statistics, a plot of [$\alpha$/Fe] versus [Fe/H] reveals a knee. The knee position, [Fe/H]$_{knee}$, corresponds to 0.1 -- 1~Gyr after the first star formation episode in a galaxy \citep[for example,][]{2020ApJ...900..179K}. 

%Despite a poor stellar statistics, results obtained by \citet{2001ApJ...548..592S} for the Sextans dSph can be considered as indicating the onset of Fe production in SNe~Ia when metal enrichment achieved [Fe/H] $\sim -2$. 
% Honda+2011 does not discuss alpha-elements

% is one of the in which the individual red giant branch stars were observed at high spectral resolution and  \citep{2001ApJ...548..592S,2009A&A...502..569A,2011PASJ...63S.523H,2010A&A...524A..58T,2020A&A...636A.111A,2020A&A...644A..75L}.
%$alpha$/Fe reflects the timescale of the chemical evolution of the system (e.g., Tinsley 1979).
%Sextans: \citet{2001ApJ...548..592S}, \citet{2009A&A...502..569A}, \citet{2011PASJ...63S.523H}, \citet{2010A&A...524A..58T}, \citet{2020A&A...636A.111A}, \citet{2020A&A...644A..75L}. 

Several studies in the literature report on contradictory results as to the trend of $\alpha$/Fe in the Sextans galaxy. With the exception of \citet{2001ApJ...548..592S}, these works were based on the DART (Dwarf Abundances and Radial velocity Team) ESO large program and the follow-up of the very metal-poor (VMP, [Fe/H] $< -2$) candidates identified in \citet{2011MNRAS.411.1013B} thanks to the DART Ca\ii\ triplet survey of Sextans.

 Investigating the first six VMP candidates, with [Fe/H] $< -2.6$, \citet[][hereafter, AAS09]{2009A&A...502..569A} found five of their targets to have solar [$\alpha$/Fe]. This can be interpreted as the possibility that the first SNe~Ia in Sextans
appeared around [Fe/H] $\sim -3$.  The only star of the sample with an overabundance of the $\alpha$-process elements, S~15-19, was reanalysed by \citet{2011PASJ...63S.523H} with the aim of understanding the origin of its strong enhancement in barium, which they concluded was due to an s-element enrichment through transfer of the processed material from an AGB.

\citet[][hereafter, AAF20]{2020A&A...636A.111A} revised atmospheric parameters of the three stars from \citet{2001ApJ...548..592S} and AAS09 using new spectra observed with HDS\footnote{High Dispersion Spectrograph}/Subaru. 
 Their analysis included the
%Their stellar sample was complemented by 
two stars from \citet{2010A&A...524A..58T} with
% Based on the newly observed HDS\footnote{High Dispersion Spectrograph}/Subaru spectra of three stars from samples of \citet{2001ApJ...548..592S} and AAS09 and 
the UVES\footnote{Ultraviolet and Visual Echelle Spectrograph}/VLT archive spectra available. 
 AAF20 confirmed their initial finding that the [$\alpha$/Fe] abundance ratios span the low
tail of the Galactic halo distribution.
%Aoki+2020: 2 stars from Aoki+2009, 1 star from Shetrone2001, 2 stars from Tafelmeyer+2010; focus on Sr/Ba \citet{2020A&A...636A.111A} 
%AAF20 infer that 'The distribution of $\alpha$/Fe abundance ratios of the Sextans dwarf galaxy stars is slightly lower than the average of the values of stars in the Galactic halo.'

\citet[][hereafter, LLP20]{2020A&A...644A..75L} presented the analysis of two new [Fe/H] $\sim -3$ stars observed with UVES/VLT. Meanwhile they reinvestigated the sample of AAS09, concluding to the existence of a normal plateau at [$\alpha$/Fe] $\sim$ 0.4. They demonstrated that previous conclusions were possibly the consequence of some bias introduced by the use of strong Mg\ione b lines.

 With a much larger metallicity coverage ($-3.2 <$ [Fe/H] $< -1.5$) and higher statistics (87 stars), \citet[][hereafter, TJL20]{2020AA...642A.176T} obtained the elemental abundances based on the high-resolution ($R \sim$ 20\,000) spectroscopy. 
%\citet[][hereafter, TJL20]{2020AA...642A.176T} obtain the elemental abundances based on the high-resolution ($R \sim$ 20\,000) spectroscopy for the largest sample of 87 stars in the Sextans dSph, which cover the wide metallicity range, $-3.2 <$ [Fe/H] $< -1.5$.
% with  taken with the multi-fibre spectrograph FLAMES/GIRAFFE. 
They find a plateau at [$\alpha$/Fe] $\sim$ 0.4 followed by decrease and estimate the location of the knee in Sextans at [Fe/H] $\sim -2$.

A homogeneous analysis of 46 stars in the Sextans galaxy performed by \citet[][hereafter, RHH20]{2020A&A...641A.127R} reveals the two knees in the [Mg/Fe] versus [Fe/H] daigram, at [Fe/H] $\sim -2.5$ and [Fe/H] $\sim -2.0$.

All together, such results make difficult to understand how star formation and chemical enrichment proceeded in the Sextans galaxy. This motivated us to perform a new homogeneous abundance analysis of the Sextans VMP stars with high-resolution spectra available, in total, 11 stars. In our previous studies of the dSphs, we demonstrated that determining a homogeneous set of atmospheric parameters and chemical abundances based on the non-local thermodynamic equilibrium (NLTE) line formation reduces an abundance scatter in individual galaxies to the point where the trends of the stellar abundance ratios with metallicity can be robustly discussed and compared between galaxies \citep{dsph_parameters,2017A&A...608A..89M,2019AstL...45..259P,2021_coma}. Here, we apply the same approach.

The paper is organised as follows. Section~\ref{Sect:obs} presents stellar sample and used observational material. Determinations of effective temperatures and surface gravities are described in Sect.~\ref{Sect:atm_param}. In Sect.~\ref{sect:abundances}, we derive metallicities and elemental abundances usung the methods described briefly in Sect.~\ref{Sect:nlte}. The obtained abundance trends are discussed in Sect.~\ref{sect:trends}. At each step of abundance analysis, comparisons with the previous high-resolution studies of the Sextans galaxy are conducted (Sects.~\ref{sect:ew}, \ref{sect:teff}, \ref{Sect:others}).
Section~\ref{sect:Conclusions} summarises our results.

\section{Stellar sample and observational material}\label{Sect:obs}

 Two Sextans stars, namely, S~11-04 and S~24-72, were already subject to NLTE analysis in our previous papers \citep[][hereafter, Paper~I and Paper~II, respectively]{dsph_parameters,2017A&A...608A..89M}. Here, we study nine new stars. Seven stars with the HDS/Subaru spectra available were selected from AAS09 and AAF20 and two stars with the UVES/VLT spectra from LLP20. 
 The UVES/VLT reduced spectra were taken from the ESO Science Archive Facility\footnote{http://archive.eso.org/cms.html}. They have $R \sim$ 34\,000 to 40\,000.  The sample stars and characteristics of the used stellar spectra are listed in Table~\ref{tab:stars}.

\begin{table*}
	\caption{Stellar sample and characteristics of the used observational material.}
	\label{tab:stars}
	\centering
	\begin{tabular}{cccll}
\hline\hline \noalign{\smallskip}
Spectral & Total exposure & S/N$^1$ & Dates  & Telescope/spectrograph,   \\
  range	[nm]          & time [h]           &     &        & PIDs  \\
\hline\hline \noalign{\smallskip}
\multicolumn{5}{l}{ \ \ \ \ \ \textbf{S~04-130}: \; RA = 10:14:28.02, \; Dec = $-$01:14:35.8, \; V = 18.07}  \\
 380-680 & 5 & 45$^2$ & 2014-12-24 & UVES/VLT, 093.D-0311(B) \\
\multicolumn{5}{l}{ \ \ \ \ \ \textbf{S~11-97}: \; RA = 10:12:27.89, \; Dec = $-$01:48:05.2, \;  V = 18.19}  \\
 380-680 & 5 & 52$^2$ & 2014-12-27 & UVES/VLT, 093.D-0311(B) \\
\multicolumn{5}{l}{ \ \ \ \ \ \textbf{S~11-04}: \; RA = 10:13:42, \; Dec = $-$02:11:24, \;  V = 17.23}  \\
 380-680 & 3.8 & 40$^2$ & 2008-04-03, 2008-04-05, 2008-04-11 & UVES/VLT, 081.B-0620A \\
\multicolumn{5}{l}{ \ \ \ \ \ \textbf{S~24-72}: \; RA = 10:15:03, \; Dec = $-$01:29:55, \;  V = 17.35}  \\
 380-680 & 4.0 & 40$^2$ & 2008-04-03, 2008-04-11 & UVES/VLT, 081.B-0620A \\
\multicolumn{5}{l}{ \ \ \ \ \ \textbf{S~10-14}: \; RA = 10:13:34.70, \; Dec = $-$02:07:57.9, \; V = 17.64}  \\
		441-565 & 6.0 & 17 - 34 & 2007-01-25, 2007-01-26, 2007-01-27 & Subaru/HDS, AAS09 \\ 
		585-704 & 6.0 & 37 - 44 & idem & idem \\ 
		394-471 & 4.3 &  8 - 16 & 2016-04-26 & Subaru/HDS, AAF20 \\ 
		485-559 & 4.3 & 16 - 23 & idem & idem  \\ 
\multicolumn{5}{l}{ \ \ \ \ \ \textbf{S~11-13}: \; RA = 10:11:42.96, \; Dec = $-$02:03:50.4, \; V = 17.53} \\
		442-565 & 3.0 &  9 - 20 & 2006-01-20 & Subaru/HDS, AAS09 \\
		586-705 & 3.0 & 22 - 27 & idem & idem \\
		585-704 & 1.0 & 21 - 24 & 2007-01-27 & idem  \\
		441-565 & 1.0 &  9 - 18 & idem & idem \\
		394-471 & 4.0 &  8 - 17 & 2016-04-27 & Subaru/HDS, AAF20 \\
		485-559 & 4.0 & 18 - 25 & idem & idem \\
\multicolumn{5}{l}{ \ \ \ \ \ \textbf{S~11-37}: \; RA = 10:13:45.48, \; Dec = $-$01:56:16.3, \; V = 17.96} \\
		441-565 & 4.0 & 14 - 28 & 2007-01-27 & Subaru/HDS, AAS09 \\
		585-704 & 4.0 & 31 - 35 & idem & idem \\
\multicolumn{5}{l}{ \ \ \ \ \ \textbf{S~12-28}: \; RA = 10:11:17.15, \; Dec = $-$02:00:24.0, \; V = 17.52}  \\
		441-565 & 4.0 & 16 - 34 & 2007-01-25 & Subaru/HDS, AAS09 \\
		585-704 & 4.0 & 39 - 45 & idem & idem  \\
\multicolumn{5}{l}{ \ \ \ \ \ \textbf{S~14-98}: \; RA = 10:13:24.48, \; Dec = $-$02:12:03.5, \; V = 18.06}  \\
		441-565 & 4.9 & 10 - 21 & 2007-01-27, 2007-01-28 & Subaru/HDS, AAS09  \\
		585-704 & 4.9 & 26 - 29 & idem & idem \\
\multicolumn{5}{l}{ \ \ \ \ \ \textbf{S~15-19}: \; RA = 10:11:26.92, \; Dec = $-$02:05:41.7, \; V = 17.54} \\
		442-565 & 11.2 & 17 - 40 & 2005-05-20, 2005-05-21, 2005-05-22, & Subaru/HDS, AAS09 \\
                        &      &         & 2005-05-23, 2006-01-20 & \\
		586-705 & 11.2 & 46 - 56 & idem & idem \\
		379-457 &  8.0 & 10 - 21 & 2010-02-09 & Subaru/HDS, HAA11 \\
		469-544 &  8.0 & 25 - 34 & idem & idem \\
\multicolumn{5}{l}{ \ \ \ \ \ \textbf{S~49}: \; RA = 10:13:11.55, \; Dec = $-$01:43:01.8, \; V = 17.52} \vspace{1mm}  \\
	
		394-471 & 4.0 &  9 - 17 & 2016-04-28 & Subaru/HDS, AAF20 \\
		485-559 & 4.0 & 17 - 24 & idem & idem  \\
\noalign{\smallskip}\hline \noalign{\smallskip}
\multicolumn{5}{l}{{\bf Notes.} $^1$ S/N is given for the reduced spectra, with noise of the sky taken into account; } \\
\multicolumn{5}{l}{$^2$ S/N refers to the 478-575~nm spectral range. } \\
\multicolumn{5}{l}{AAS09 = \citet{2009A&A...502..569A}, AAF20 = \citet{2020A&A...636A.111A}, HAA11 = \citet{2011PASJ...63S.523H}.} \\
\noalign{\smallskip} \hline
	\end{tabular}
\end{table*} 

\subsection{Reduction of the Subaru spectra}

For seven stars, we used the HDS/Subaru spectra observed with $R \sim$ 40\,000. The raw data were taken from the Subaru-Mitaka-Okayama-Kiso Archive (SMOKA)\footnote{\url{https://smoka.nao.ac.jp/}} \citep{2002ASPC..281..298B}. Standard data reduction procedures including bias, cosmic ray strikes, sky and background scattered light subtraction, wavelength calibration, spectral extraction, and order merging were carried out with the MIDAS\footnote{Munich Image Data Analysis System} \textit{echelle} package. 

Removing the sky light is important for faint objects, such as our sample stars of 17 to 18 visual magnitudes. In such a case, the signal levels on the CCD are comparable for the star and the night sky. Therefore, one needs to determine correctly a position of the star in the slit and a contribution of the sky background to the object's light. The MIDAS standard procedure uses slice along the slit. For noisy spectrum, this does not work well enough. Lower signal-to-noise ratio (S/N) of the used HDS/Subaru spectra compared with other similar high-resolution data at similar magnitudes is, in part, due to observing close to the full moon (AAS09, AAF20) and shorter exposure time (see Table~\ref{tab:stars}). To increase the signal at a given position on the slit, we averaged pixels along the corresponding spectral order on the CCD. Using 35 echelle orders composed of 2050 pixels each, we obtained the profile that includes both the star and the sky signals, see Fig.~\ref{fig:profile} for the star S~15-19 observed on 2005-05-20 with an exposure time of 1\,800~s.
%shows profiles of the CCD counts along the slit for  It was obtained by summing 71\,750 pixels (35 spectral echelle orders composed of 2050 pixels each) on the CCD frame. 
The profile produced by the star was approximated by a Gaussian, and for the sky background we used a more complex function, namely, Gaussians on the borders and a constant value in the center of the CCD frame. Such a procedure made it possible to fix a position of the star on the slit and to determine the ranges for calculations of signals from the star and the sky. 

To remove the cosmic ray traces, we applied a two-step procedure.
At the first step, using the MIDAS standard procedure, we created a median averaged frame from the observational frames. This is enough in most cases, but does not work well for noisy
spectra. At the second step, each pixel on each observed frame was checked with the corresponding
pixel on the median averaged frame.
If the pixel signal exceeds that for the averaged frame by more than four times, it was marked as
cosmic one and replaced with the value from the averaged frame.
Finally, all the corrected observational frames were summed up. 

Individual \textit{echelle} orders were merged after normalisation of the spectra. The wavelength scale was corrected by accounting for the Earth motion.

\subsection{Comparisons with the literature measurements}\label{sect:ew}

\begin{figure*}  %[htbp]
 \begin{minipage}{170mm}
\centering
	\includegraphics[width=0.99\textwidth, clip]{EW_cmp.eps}
  \caption{\label{Fig:subaru} Equivalent widths determined in this study, EW(ts), compared with measurements of \citet[][LLP20]{2020A&A...644A..75L}, \citet[][HAA11]{2011PASJ...63S.523H}, \citet[][AAF20]{2020A&A...636A.111A}, and \citet[][AAS09]{2009A&A...502..569A} for the selected Sextans stars. The regression lines are shown by continuous lines.
}
\end{minipage}
\end{figure*}

\begin{figure}  %[htbp]
 \begin{minipage}{85mm}
\centering
	\includegraphics[width=0.99\textwidth, clip]{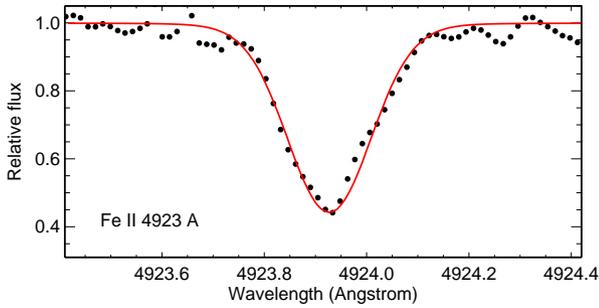}
  \caption{\label{Fig:fe4923} Best fit (continuous curve) to the Fe\ii\ 4923\,\AA\ line in S~11-97 (filled circles). We measured EW = 112.6\,m\AA. }
\end{minipage}
\end{figure}

 Despite that elemental abundances were derived in this study via the synthetic spectrum method (see Sect.~\ref{Sect:codes}), we measured also equivalent widths (EWs) of the lines  with a purpose of comparison with the literature data based on the same observations. We used the code {\sc synthV}\_NLTE \citep{2019ASPC} and a Voigt profile for reproducing the observed line profile. 
%the best fits to the observed spectra. In Fig.~\ref{Fig:subaru}, these EWs are compared with the literature measurements based on the same observations. For the Subaru spectra, 
The EW errors were estimated from deviations of the theoretical spectrum, $y_i^{th}$, from the observed one, $y_i$:
$$
\sigma_{EW} = \sqrt{\frac{\sum\limits_{i=1}^{n}(y_i - y_i^{th})^2}{n-1}}
\Delta\lambda
$$

\noindent
Here, $n$ is a total number of wavelength points over the line profile and $\Delta\lambda$ is the total width of the line.  Equivalent width comparisons are presented in Fig.~\ref{Fig:subaru}.

For S~04-130 and S~11-97, with the UVES/VLT spectra available, our EWs are fairly consistent with that reported by LLP20. The exceptions are 
the strongest lines of Fe\ii, at 4923 and 5018~\AA. For S~11-97, LLP20 measure EW = 77.4 and 82.2~m\AA, respectively, while our values are 112.6~m\AA\ (see Fig.~\ref{Fig:fe4923} for a justification) and 120.1~m\AA. 

Only minor discrepancies were found between our measurements and those of AAS09 (for example, S~11-37) and AAF20 (for example, S~49). 
  For S~15-19, our EWs are slightly smaller compared to those of \citet[][hereafter, HAA11]{2011PASJ...63S.523H}. This can be due to overestimation of EWs in HAA11. Indeed, the latter paper reports that that their measured EWs are larger than those of AAS09, by 6\%, whereas our measurements agree well with those of AAS09. 

For the Subaru stars, we determined their radial heliocentric velocities $V_{rad}$.
They are listed in Table~\ref{Tab:parameters} and agree within the error bars with measurements of AAS09. 

\begin{figure}  %[htbp]
 \begin{minipage}{85mm}
\centering
	\includegraphics[width=0.99\textwidth, clip]{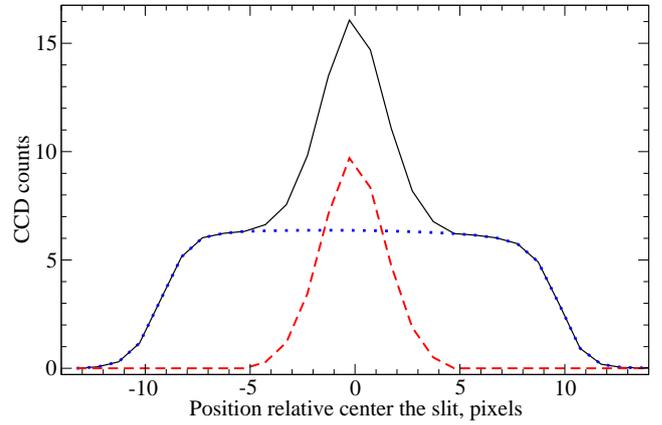}
  \caption{\label{fig:profile} Profiles of the CCD counts along the slit. The continuous curve corresponds to the average over the spectrum at a given X coordinate. The recovered sky light is shown by dotted curve and the extracted stellar profile by dashed curve. The data corresponds to the star S~15-19 in the 2005-05-20 night. }
\end{minipage}
\end{figure}

\begin{table*}
	\caption{Sources of photometric data for spectral energy distributions of the sample stars}\label{tab:obs}
	\centering
	%\tabcolsep=4pt
%	\renewcommand{\arraystretch}{1.0}
%\small
\begin{tabular}{rlcccccccccccll}	
\hline 
%$\lambda$, \AA & Filter & \multicolumn{11}{c}{Stars} & Vizier catalog & Reference\ \\
%\cline{3-13} \\
$\lambda$, \AA & Filter & 
\begin{sideways}S04-130\end{sideways} &
\begin{sideways}S10-14\end{sideways} &
\begin{sideways}S11-04\end{sideways} &
\begin{sideways}S11-13\end{sideways} &
\begin{sideways}S11-37\end{sideways} &
\begin{sideways}S11-97\end{sideways} &
\begin{sideways}S12-28\end{sideways} &
\begin{sideways}S14-98\end{sideways} &
\begin{sideways}S15-19\end{sideways} &
\begin{sideways}S24-72\end{sideways} &
\begin{sideways}S49\end{sideways} & Vizier catalog & Reference \\
\hline 
3519 & SDSS:u &  - & - & + & + & - & - & + & + & - & + & - &  II/296/catalog  &  \cite{vizier:II/296} \\
3519 & SDSS:u &  + & + & + & + & + & + & + & + & + & + & + &  IV/38/tic  &  \cite{vizier:IV/38} \\
3519 & SDSS:u &  - & - & - & - & - & + & - & - & - & - & - &  J/AJ/151/41/movers  &  \cite{vizier:J/AJ/151/41} \\
4442 & Johnson:B &  + & + & + & + & + & - & + & - & - & + & + &  I/305/out  &  \cite{vizier:I/305} \\
4442 & Johnson:B &  + & + & + & + & + & - & + & - & - & + & + &  IV/38/tic  &  \cite{vizier:IV/38} \\
4442 & Johnson:B &  + & - & - & - & - & - & - & - & - & - & - &  J/A+A/581/A41/tablea2  &  \cite{vizier:J/A+A/581/A41} \\
4680 & POSS-II:J &  + & + & + & + & + & + & - & - & - & + & + &  I/305/out  &  \cite{vizier:I/305} \\
4772 & PAN-STARRS:g &  - & - & - & - & + & - & + & + & - & - & - &  I/343/gps1  &  \cite{vizier:I/343} \\
4772 & PAN-STARRS:g &  + & + & + & + & + & + & + & + & + & + & + &  II/349/ps1  &  \cite{vizier:II/349} \\
4819 & SDSS:g &  - & - & + & + & - & - & + & - & - & + & - &  II/296/catalog  &  \cite{vizier:II/296} \\
4819 & SDSS:g &  + & + & + & + & + & + & + & + & + & + & + &  IV/38/tic  &  \cite{vizier:IV/38} \\
4819 & SDSS:g &  + & + & - & + & + & + & + & + & + & - & + &  J/A+A/609/A53/table2  &  \cite{vizier:J/A+A/609/A53} \\
4968 & SkyMapper:g &  - & - & + & - & - & - & - & - & - & + & - &  II/358/smss  &  \cite{vizier:II/358} \\
5035 & GAIA3:Gbp &  - & - & + & - & - & - & - & - & - & + & - &  I/350/gaiaedr3  &  \cite{vizier:I/350} \\
5046 & GAIA2:Gbp &  + & + & + & + & + & + & + & + & + & + & + &  I/345/gaia2  &  \cite{2018AA...616A...1G} \\
5537 & Johnson:V &  + & + & + & + & + & + & + & + & + & + & + &  IV/38/tic  &  \cite{vizier:IV/38} \\
5537 & Johnson:V &  + & - & - & - & - & - & - & - & - & - & - &  J/A+A/581/A41/tablea2  &  \cite{vizier:J/A+A/581/A41} \\
5537 & Johnson:V &  - & - & - & - & - & - & - & - & - & - & + &  J/AJ/137/3100/stars  &  \cite{vizier:J/AJ/137/3100} \\
5822 & GAIA3:G &  - & - & + & - & - & - & - & - & - & + & - &  I/350/gaiaedr3  &  \cite{vizier:I/350} \\
6040 & SkyMapper:r &  - & - & + & - & - & - & - & - & - & + & - &  II/358/smss  &  \cite{vizier:II/358} \\
6125 & PAN-STARRS:r &  + & + & + & + & + & + & + & + & + & + & + &  II/349/ps1  &  \cite{vizier:II/349} \\
6226 & GAIA2:G &  + & + & - & + & - & - & - & - & + & + & + &  I/345/gaia2  &  \cite{2018AA...616A...1G} \\
6246 & SDSS:r &  - & - & + & + & - & - & + & + & - & + & - &  II/296/catalog  &  \cite{vizier:II/296} \\
6246 & SDSS:r &  + & + & + & + & + & + & + & + & + & + & + &  IV/38/tic  &  \cite{vizier:IV/38} \\
6246 & SDSS:r &  + & + & - & - & + & + & + & + & + & - & + &  J/A+A/609/A53/table2  &  \cite{vizier:J/A+A/609/A53} \\
6246 & SDSS:r &  + & + & + & + & + & + & + & - & + & + & + &  J/AJ/141/189/table4  &  \cite{vizier:J/AJ/141/189} \\
6246 & SDSS:r &  - & - & - & - & - & + & - & - & - & - & - &  J/AJ/151/41/movers  &  \cite{vizier:J/AJ/151/41} \\
6399 & POSS-II:F &  + & + & + & + & + & + & - & + & - & - & + &  I/305/out  &  \cite{vizier:I/305} \\
7479 & PAN-STARRS:i &  + & - & + & + & + & + & + & + & + & + & + &  II/349/ps1  &  \cite{vizier:II/349} \\
7620 & GAIA3:Grp &  - & - & + & - & - & - & - & - & - & + & - &  I/350/gaiaedr3  &  \cite{vizier:I/350} \\
7634 & SDSS:i &  - & - & + & + & - & - & + & + & - & + & - &  II/296/catalog  &  \cite{vizier:II/296} \\
7634 & SDSS:i &  + & + & + & + & + & + & + & + & + & + & + &  IV/38/tic  &  \cite{vizier:IV/38} \\
7634 & SDSS:i &  - & - & - & - & - & + & - & - & - & - & - &  J/AJ/151/41/movers  &  \cite{vizier:J/AJ/151/41} \\
7724 & GAIA2:Grp &  + & + & + & + & + & + & + & + & + & + & + &  I/345/gaia2  &  \citet{2018AA...616A...1G} \\
7836 & POSS-II:i &  + & - & + & + & - & + & + & + & - & + & + &  I/305/out  &  \cite{vizier:I/305} \\
8652 & PAN-STARRS:z &  - & + & + & + & + & + & + & + & + & + & + &  II/349/ps1  &  \cite{vizier:II/349} \\
9017 & SDSS:z &  - & - & + & + & - & - & + & + & - & + & - &  II/296/catalog  &  \cite{vizier:II/296} \\
9017 & SDSS:z &  + & + & + & + & + & + & + & + & + & + & + &  IV/38/tic  &  \cite{vizier:IV/38} \\
9596 & PAN-STARRS:y &  + & + & + & + & + & + & + & + & + & + & + &  II/349/ps1  &  \cite{vizier:II/349} \\
10304 & UKIDSS:Y &  + & + & + & + & + & + & + & + & + & + & + &  II/319/las9  &  \cite{vizier:II/319} \\
12390 & 2MASS:J &  - & - & - & - & - & - & - & - & + & - & - &  I/297/out  &  \cite{vizier:I/297} \\
12390 & 2MASS:J &  - & - & - & - & - & - & - & - & + & - & - &  I/327/cmc15  &  \cite{vizier:I/327} \\
12390 & 2MASS:J &  + & + & + & + & + & + & + & + & - & + & + &  II/311/wise  &  \cite{vizier:II/311} \\
12483 & UKIDSS:J &  + & + & + & + & + & + & + & + & + & + & + &  II/319/las9  &  \cite{vizier:II/319} \\
12489 & UKIRT/WFCAM:J &  - & - & - & - & - & + & - & - & - & - & - &  J/A+A/642/A176/table2  &  \citet{2020AA...642A.176T} \\
12500 & Johnson:J &  + & + & + & + & + & - & + & + & - & + & + &  II/246/out  &  \cite{vizier:II/246} \\
12500 & Johnson:J &  + & + & + & + & + & + & + & + & + & + & + &  IV/38/tic  &  \cite{vizier:IV/38} \\
16300 & Johnson:H &  + & + & + & + & + & + & - & + & - & + & + &  II/246/out  &  \cite{vizier:II/246} \\
16300 & Johnson:H &  + & + & + & + & + & + & + & + & + & + & + &  IV/38/tic  &  \cite{vizier:IV/38} \\
16312 & UKIDSS:H &  + & + & + & + & + & + & + & + & - & + & + &  II/319/las9  &  \cite{vizier:II/319} \\
16338 & UKIRT/WFCAM:H &  - & - & - & - & - & + & - & - & - & - & - &  J/A+A/642/A176/table2  &  \cite{2020AA...642A.176T} \\
16494 & 2MASS:H &  - & - & - & - & + & - & - & - & + & - & - &  I/297/out  &  \cite{vizier:I/297} \\
16494 & 2MASS:H &  - & - & - & - & - & - & - & - & + & - & - &  I/327/cmc15  &  \cite{vizier:I/327} \\
16494 & 2MASS:H &  - & - & - & + & - & - & - & - & - & - & - &  I/339/hsoy  &  \cite{vizier:I/339} \\
16494 & 2MASS:H &  + & + & + & + & + & + & + & + & - & + & + &  II/311/wise  &  \cite{vizier:II/311} \\
21637 & 2MASS:Ks &  - & - & - & - & + & - & - & - & + & - & - &  I/297/out  &  \cite{vizier:I/297} \\
21637 & 2MASS:Ks &  + & + & - & + & + & + & + & - & - & - & + &  II/311/wise  &  \cite{vizier:II/311} \\
21637 & 2MASS:Ks &  - & - & + & - & - & - & - & - & + & + & - &  IV/38/tic  &  \cite{vizier:IV/38} \\
21900 & Johnson:K &  - & + & + & + & + & - & + & - & + & + & + &  II/246/out  &  \cite{vizier:II/246} \\
22009 & UKIDSS:K &  + & + & + & + & + & + & + & + & + & + & + &  II/319/las9  &  \cite{vizier:II/319} \\
\hline\hline
\end{tabular}
\end{table*} 

\begin{table*}
	\contcaption{ -- Sources of photometric data for spectral energy distributions of the sample stars}\label{tab:continue}
	\centering
\begin{tabular}{rlcccccccccccll}	
	\hline 
%	$\lambda$, \AA & Filter & \multicolumn{11}{c}{Stars} & Vizier catalog & Reference\ \\
%	\cline{3-13} \\
$\lambda$, \AA 	& Filter & 
	\begin{sideways}S04-130\end{sideways} &
	\begin{sideways}S10-14\end{sideways} &
	\begin{sideways}S11-04\end{sideways} &
	\begin{sideways}S11-13\end{sideways} &
	\begin{sideways}S11-37\end{sideways} &
	\begin{sideways}S11-97\end{sideways} &
	\begin{sideways}S12-28\end{sideways} &
	\begin{sideways}S14-98\end{sideways} &
	\begin{sideways}S15-19\end{sideways} &
	\begin{sideways}S24-72\end{sideways} &
	\begin{sideways}S49\end{sideways} & 
	& \\
	\hline 
22185 & UKIRT/WFCAM:K &  - & - & - & - & - & + & - & - & - & - & - &  J/A+A/642/A176/table2  &  \cite{2020AA...642A.176T} \\
33500 & WISE:W1 &  + & + & + & + & + & + & + & + & - & + & + &  II/311/wise  &  \cite{vizier:II/311} \\
33500 & WISE:W1 &  + & - & - & + & - & - & - & - & - & - & + &  II/328/allwise  &  \cite{vizier:II/328} \\
33500 & WISE:W1 &  + & + & - & + & + & - & - & - & + & + & + &  II/363/unwise  &  \cite{vizier:II/363} \\
33500 & WISE:W1 &  - & - & + & - & - & - & - & - & - & + & - &  II/365/catwise  &  \cite{vizier:II/365} \\
33500 & WISE:W1 &  - & + & + & + & + & + & - & + & - & + & + &  IV/38/tic  &  \cite{vizier:IV/38} \\
46000 & WISE:W2 &  + & + & + & + & + & + & + & + & - & + & - &  II/311/wise  &  \cite{vizier:II/311} \\
46000 & WISE:W2 &  - & - & + & + & - & - & - & - & - & - & + &  II/328/allwise  &  \cite{vizier:II/328} \\
46000 & WISE:W2 &  - & + & - & + & + & + & - & - & + & + & + &  II/363/unwise  &  \cite{vizier:II/363} \\
46000 & WISE:W2 &  - & - & - & - & - & + & - & - & - & - & - &  III/284/allvis  &  \cite{vizier:III/284} \\
46000 & WISE:W2 &  - & + & - & + & + & - & + & + & - & + & - &  IV/38/tic  &  \cite{vizier:IV/38} \\	
\hline\hline
\end{tabular}
\end{table*} 

%\section{Atmospheric parameters}\label{Sect:atm_param}

\section{Effective temperatures and surface gravities}\label{Sect:atm_param}

\begin{figure}  %[htbp]
 \begin{minipage}{85mm}
\centering
	\includegraphics[width=0.99\textwidth, clip]{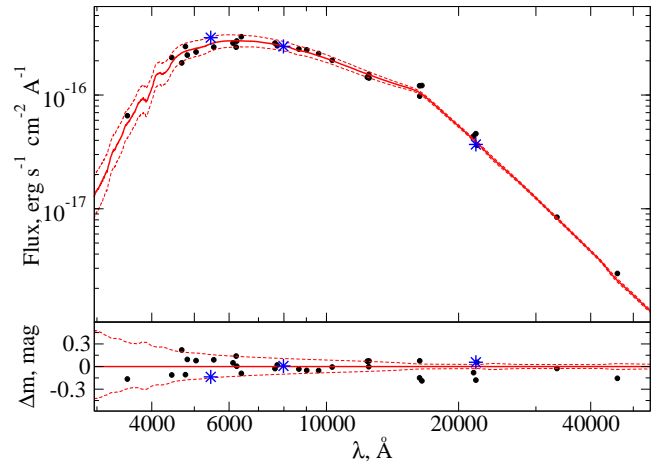}
  \caption{\label{fig:SED} Best fit (continuous curve, $\Teff$ = 4590~K, $\logg$ = 1.01, [Fe/H] = $-2.6$) to the spectral energy distribution based on the Vizier data for the star S~10-14 (circles). Fluxes of the $VIK$ bands from \citet{2009A&A...502..569A} are shown by asterisks. Dashed curves show the theoretical fluxes for 100~K higher and lower temperatures. The differences between observed and theoretical fluxes (in magnitudes) are displayed in the bottom part.}
\end{minipage}
\end{figure}

For each sample star, its $\Teff$ and $\logg$ were derived from the spectral energy distribution (SED) that was constructed using the calibrated fluxes provided by the 
%This approach we used due to lack of accurate measurements for visual and IR fluxes of studied stars.  multicolor photometry The calibrated data was obtained using 
VizieR Photometry viewer service \footnote{\url{http://vizier.u-strasbg.fr/vizier/sed/}}.  The Vizier service uses many photometric catalogues. Some of them were obtained by compiling the literature data and include common sources. However, different catalogues can indicate different fluxes, when citing the same original source of photometric data. Discrepancies can originate from using different data releases (for example, based on the SDSS data) and/or different transformation of the photometric magnitude to the flux. We did not employ the sources with clearly outlying fluxes, by more than 30~\%\ compared to the fluxes from the other sources. The used sources of the original photometric data are listed in Table~\ref{tab:obs}. In case of multiple sources for one wavelength, we computed the average flux. The data is available in the 3519-220\,906~\AA\ wavelength range, however, we used the $\lambda \le$ 50\,000~\AA\ range, which is less affected by the interstellar dust absorption. The observed ($f^{obs}_{\lambda}$) and emergent ($F_{\lambda}$), stellar fluxes are related, as follows: 
\begin{equation}
 f^{obs}_{\lambda} = F_{\lambda}(\Teff,\logg,[Fe/H]) \times f_{\lambda}(\Ebv) \frac{\theta^2}{4} .
\label{formula:flux}
\end{equation}

\noindent
Here, $f_{\lambda}(\Ebv)$ is the interstellar absorption function according to \citet{1990ARA&A..28...37M} and $\theta$ is an angular diameter. Following AAS09, we adopted the foreground reddening of $\Ebv$ = 0.038, which was estimated from the dust maps of \citet{1998ApJ...500..525S}.

For $F_{\lambda}(\Teff,\logg,[Fe/H])$, we used the theoretical fluxes of \citet{Gustafssonetal:2008}, as available at the MARCS\footnote{\url{https://marcs.astro.uu.se}} website. They were convolved with a Gaussian of 300-500~\AA\ full width at half maximum (FWHM) for the ultra-violet and visual wavelengths and FWHM = 1000~\AA\ for the infrared (IR) wavelengths.

The effective temperature and $\logg$ were evaluated in the course of successive iterations. Metallicity was fixed at [Fe/H] = $-2.7$ because the theoretical fluxes of cool VMP giants are only weakly sensitive to variations in [Fe/H]. First, we calculated $\theta = 2 \sqrt{f_{\rm IR}/(F_{\rm IR}\times f_{IR}(\Ebv))}$ using the IR fluxes, which depend weakly on $\Teff$ variations. For initial values, we adopted $\Teff$ = 4500~K and $\logg$ = 1.2. Next, $\Teff$ was improved from fitting the theoretical SED to the observed one, as shown in Fig.~\ref{fig:SED}. In the fitting procedure, we applied the Levenberg-Marquardt minimization method \citep{Levenberg_1944,Marquardt_1963}. Then we calculated the surface gravity:
\begin{equation}
\logg = 4.44 + \log {\it M/M}_\odot - 2 \log {\it R/R}_\odot ,
\label{formula:logg}
\end{equation}

\noindent
assuming a star's mass of 0.8$M_\odot$ and computing the star's radius as
$ R/R_\odot = 1.0752 \cdot 10^{-4} \times \theta \times d $ 
with $d = 90\pm10$~kpc \citep{2013AJ....145..101K} and $\theta$ in $\mu$as.
The procedure was repeated until the results converge. 

The obtained stellar parameters are presented in Table~\ref{Tab:parameters}. 
The uncertainty in $\logg$ is mainly determined by the distance error and, for the Sextans stars, amounts to approximately 0.1~dex. The uncertainty in the SED fitting results in $\Teff$ errors of 20--30~K, which are defined as a square root of the dispersion value from the covariation matrix. However, from the differences between observed and theoretical fluxes we estimated the uncertainty in $\Teff$ as 100~K.

\begin{table*} % [htbp]
 \caption{\label{Tab:parameters} Atmospheric parameters, angular diameters, and radial heliocentric velocities of the sample stars.} 
 \centering
 \begin{tabular}{lclcccc}
\hline\hline \noalign{\smallskip}
ID & $\Teff$ &  log~$g$ &  [Fe/H] &  $\xi_t$ & $\theta$      & $V_{rad}$   \\
&   [K]      &          &         & [\kms]   & [$\mu$as]       &   [\kms] \\  
\noalign{\smallskip} \hline \noalign{\smallskip}
S~04-130  & 4610(24) & 1.15(0.11) &  $-2.72$(0.20) & 2.1  & 4.06(0.07) & 215.5(1.0)$^1$ \\
S~11-97   & 4660(10) & 1.22(0.12) &  $-2.70$(0.19) & 2.1  & 3.78(0.01) & 218.3(1.1)$^1$ \\
S~11-04$^2$ & 4380(120) &  0.57(0.10) &  $-$2.63(0.18) & 2.2 & 7.67(0.10) & - \\
S~24-72$^2$ & 4400(40) &  0.76(0.09) &  $-$2.82(0.15) & 2.2 & 6.76(0.09) & - \\
S~10-14 & 4590(25) &  1.01(0.10) &  $-$2.97(0.22) & 2.1 & 4.81(0.08) & 234.0(1.4) \\
S~11-13 & 4470(15) &  0.85(0.08) &  $-$3.05(0.18) & 2.1 & 5.76(0.06) & 224.5(1.3) \\
S~11-37 & 4700(25) &  1.15(0.11) &  $-$2.89(0.15) & 2.1 & 4.06(0.06) & 222.1(1.1) \\
S~12-28 & 4590(20) &  0.95(0.09) &  $-$2.86(0.18) & 2.1 & 5.13(0.06) & 201.8(1.1) \\
S~14-98 & 4700(20) &  1.25(0.12) &  $-$2.73(0.17) & 2.1 & 3.64(0.01) & 223.7(1.9) \\
S~15-19 & 4540(30) &  0.95(0.09) &  $-$3.32(0.16) & 2.1 & 5.13(0.09) & 225.2(1.4) \\
S~49    & 4470(20) &  0.85(0.08) &  $-$3.02(0.17) & 2.1 & 5.79(0.07) & 233.1(1.4)\\ 
\noalign{\smallskip}\hline \noalign{\smallskip}
\multicolumn{7}{l}{{\bf Notes.} The numbers in parentheses are the errors of the derived parameters.} \\
\multicolumn{7}{l}{ $^1$ from \citet{2020A&A...644A..75L}.} \\
\multicolumn{7}{l}{ $^2$ Atmospheric parameters from \citet{dsph_parameters}, except the angular diameter.} \\
\noalign{\smallskip} \hline
\end{tabular}
\end{table*}

\subsection{Comparison with the methods from Paper~I}

We tend to produce a homogeneous set of atmospheric parameters for all the stars of our project. The method we applied in this study to determine $\Teff$ and $\logg$ is more straightforward compared with that proposed in our Paper~I. We checked the differences between the two approaches.

For S~04-130 and S~11-97, their $\Teff$ were derived from the $V-K$, $V-J$, and $V-H$ colours and calibration of \citet{2005ApJ...626..465R} and $\logg$ from the known distance. The obtained $\Teff$ = 4630$\pm$77~K (S~04-130) and 4610$\pm$20~K (S~11-97) are consistent within 50~K with the temperatures determined in this study from the SED fitting. For $\logg$, the differences between the methods applied in Paper~I and this study amount to $-0.09$~dex and $-0.13$~dex, which are within the error bars of surface gravity determinations (see Table~\ref{Tab:parameters}).  

%Our sample of very metal-poor (VMP, [Fe/H] $< -2$) stars in the Sextans dSph has been and includes. 
Table~\ref{Tab:parameters} includes the stars S~11-04 and S~24-72, for which the NLTE abundance analyses were made in Paper~II using photometric $\Teff$ from \citet{2010A&A...524A..58T} and the distance-based $\logg$ from Paper~I. The differences between this study and Paper~I amount to $-40$~K and 0~K for $\Teff$ and +0.03 and $-0.05$~dex for $\logg$. For S~11-04, moving to revised $\Teff$/$\logg$ would shift abundances by 0.04-0.05~dex and up to 0.02~dex for lines of the minority (Fe\ione, Ca\ione, etc.) and the majority (Fe\ii, Ba\ii, etc.) species, respectively. For S~24-72, the corresponding numbers are 0 and 0.01~dex. Therefore, in this study, we did not revise elemental abundances of S~11-04 and S~24-72 (the exception is Sr, see Sect.~\ref{Sect:nlte}) and, in Sects.~\ref{Sect:others} and \ref{sect:trends}, use the data from Paper~II. 

\subsection{Comparisons with the literature data}\label{sect:teff}

\begin{figure}  %[htbp]
 \begin{minipage}{85mm}
\centering
	\includegraphics[width=0.99\textwidth, clip]{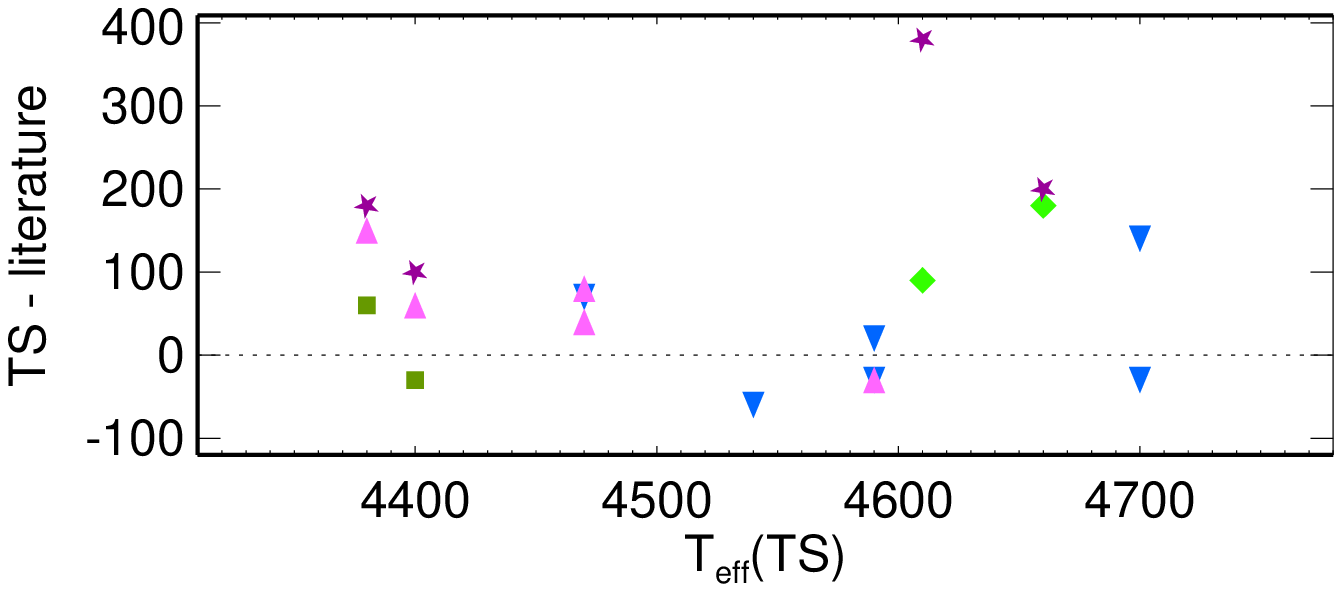}
	
\vspace{-5mm}
	\includegraphics[width=0.99\textwidth, clip]{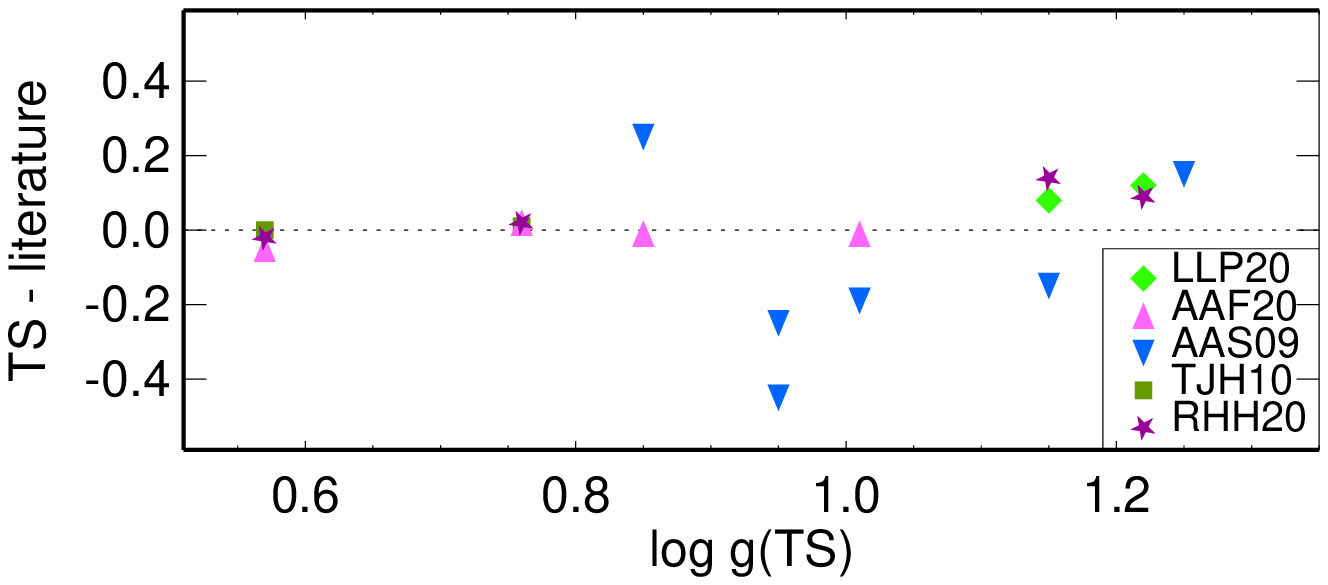}
	
%\vspace{-5mm}
%	\includegraphics[width=0.99\textwidth, clip]{compare_feh.ps}
  \caption{\label{Fig:comp1} Differences in atmospheric parameters between this study (TS) and previous studies of \citet[][AAS09, inverted triangles]{2009A&A...502..569A}, \citet[][TJH10, squares]{2010A&A...524A..58T}, \citet[][AAF20, triangles]{2020A&A...636A.111A}, \citet[][LLP20, rhombi]{2020A&A...644A..75L}, and \citet[][RHH20, five-pointed stars]{2020A&A...641A.127R}. }
\end{minipage}
\end{figure}

Effective temperatures and surface gravities determined in this study are compared in Fig.~\ref{Fig:comp1} with the literature data for common stars. 
%We adopt 100~K as a typical error of $\Teff$ determinations. 
The differences with photometric temperatures of AAS09 and AAF20 reveal a scatter between $-60$~K and 140~K. The only color, $V-K$, was used by AAS09 and AAF20 to determine $\Teff$, and they estimate the uncertainty in the effective temperatures as $\pm$150~K.
%This can be due to errors of the $V$ or/and $K$ magnitudes in AAS09 and AAF20, because only $V-K$ was used to determine $\Teff$. 
\citet{2020A&A...644A..75L} used the Fe\ione\ excitation equilibrium method that is known to underestimate effective temperatures compared with the photometric ones \citep[see][and references therein]{2013ApJ...769...57F}. For S~04-130 and S~11-97, the differences amount to 90 and 180~K, respectively. At the same time, we agree well with \citet{2010A&A...524A..58T} who applied the same method as LLP20. 
The largest and systematic discrepancies, of 215~K, on average, were found with RHH20 who derived $\Teff$ from flux ratios in the spectral region around the Balmer lines H$_\alpha$ and H$_\beta$ using the automated code. For comparison, the difference with all the remaining studies amounts to 47~K, on average.

Our $\logg$ values agree well with the data from TJH10, AAF20, LLP20, and RHH20, which are based on the Sextans dSph distance. While discrepancies with AAS09, who used the Fe\ione /Fe\ii\ ionization equilibrium method under the LTE assumption, range between $-0.45$~dex and 0.25~dex.

\section{NLTE calculations}\label{Sect:nlte}

Abundances of Na, Mg, Al, Si, Ca, Ti, Fe, Sr, Zr, and Ba were determined in this study based on the NLTE line formation, with using the model atoms and  methods developed by \citet{alexeeva_na,mash_mg13,Baumueller_al1,2020MNRAS.493.6095M,mash_ca,sitnova_ti,mash_fe,1997ARep...41..530B,Velichko2010_zr}, and \citet{Mashonkina1999}, respectively. At the time of constructing the model atoms listed, accurate data on inelastic processes in collisions with neutral hydrogen atoms (H\ione) were missing, and therefore we applied a rough theoretical approximation of \citet{Drawin1969}, as implemented by \citet{Steenbock1984}. Now, for most chemical species under investigation, we employ quantum-mechanical rate coefficients from calculations of \citet[][Na\ione]{barklem2010_na}, \citet[][Mg\ione]{mg_hyd2012}, \citet[][Al\ione]{Belyaev2013_Al}, \citet[][Si\ione]{Belyaev2014_Si}, \citet[][Ca\ione]{2017ApJ...851...59B}, and \citet[][Ba\ii]{2018MNRAS.478.3952B}. This paper presents for the first time calculations of the Sr\ii\ + H\ione\ collisions based on the quantum model approach in the framework of the Born-Oppenheimer approximation  (Sect.~\ref{sect:hyd}) and the NLTE analysis of the Sr\ii\ lines using accurate data on not only electron but also atomic hydrogen impact excitation. For Ti\ione -\ii, Fe\ione -\ii, and Zr\ii, collisions with H\ione\ were treated in this study using the Drawinian rates scaled by factors of $S_{\rm H}$ = 1, 0.5, and 0.1, respectively.

\subsection{Calculations of the Sr\ii\ + H\ione\ collisions}\label{sect:hyd}

The study of the inelastic processes in collisions of Sr\ii\ = Sr$^+$ with hydrogen is performed using the quantum model approach in the framework of the Born-Oppenheimer approximation \citep{Yakovleva:2016aa}.
The \textit{ab initio} calculations performed by \citet{Aymar_SrH+} and by \citet{Gadea_SrH+} provide the SrH$^+$ electronic structure for several lower-lying states, but do not provide the data for the fine structure levels of the quasimolecule.
For this reason, we employ the approach proposed by \citet{Belyaev:2019pra}  for collisions of alkali-like atoms and ions with hydrogen. 
This approach suggests to model the asymptotic diabatic potentials for the reaction channels ${\rm A}^{Z+}({\rm^1S_{0}}) + {\rm H}^-({\rm^1S_{0}})$ and ${\rm A}^{(Z-1)+}({\rm^2L_{j}}) + {\rm H} ({\rm^2S_{\rfrac{1}{2}}})$ and divide the off-diagonal diabatic hamiltonian matrix elements for the fine structure levels proportional to the coefficients containing the Clebsch-Gordan coefficients \citep[see][for details]{Belyaev:2019pra}.
The Clebsch-Gordan coefficients appear due to the transformation of the electronic structure from the LS representation (without the fine structure levels) to the JJ representation (with the fine structure levels).

The ionic reaction channel ${\rm Sr}^{2+}(4p^{6}~{\rm^1S_{0}}) + {\rm H}^-(1s^2~{\rm^1S_{0}})$ has the same structure as in case of alkali-hydrogen collisions because there is the only active electron and the ionic molecular state has $0^+$ molecular symmetry. 
The calculations of the SrH$^+$ electronic structure are performed for 31 scattering channels listed in Table~\ref{tab:states}. The excitation energies of 30 lowest states of Sr$^+$ and the ionizaton energy of the ground state of Sr$^{2+}$ were taken from NIST \citep{NIST_ASD}.
The non-adiabatic nuclear dynamics is investigated using the multichannel Landau-Zener approach \citep{Yakovleva:2016aa}.
The mutual neutralization, ion pair production, excitation and de-excitation inelastic processes are treated.
The cross sections and rate coefficients are calculated for all inelastic processes due to the transitions between the states listed in Table~\ref{tab:states}. 

\begin{figure}           
\includegraphics[scale=0.43]{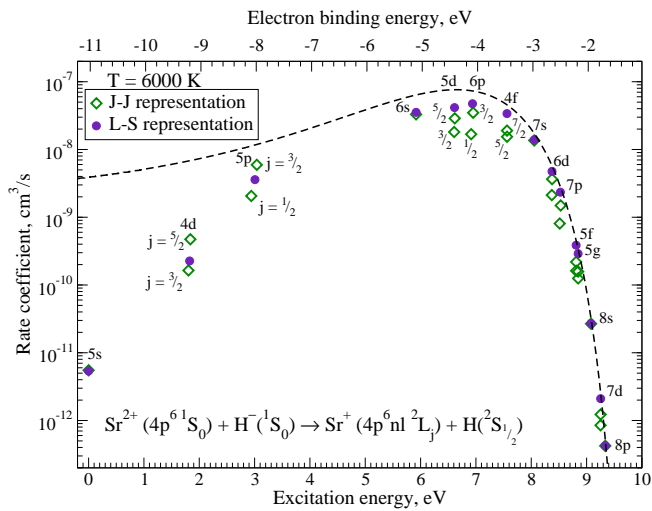}
\caption{The rate coefficients (in cm$^3$/s) for the mutual neutralization processes in ${\rm Sr}^{2+} + {\rm H}^-$ at temperature T = 6000~K as a function of the final channel excitation energy (or electron binding energy).}
\label{fig:neutr_rates}
\end{figure}

The analysis of the calculated rate coefficients shows that the largest rates correspond to the mutual neutralization processes to states 6 -- 13 (see Table~\ref{tab:states}).
Figure~\ref{fig:neutr_rates} demonstrates the mutual neutralization rate coefficients calculated both with and without the account for the fine structure levels of Sr$^+$. 
The calculations in LS and JJ representations give almost the same rate coefficients for mutual neutralization to Sr$^{+}({\rm^2S_{\rfrac{1}{2}}})$ + H states, while the results for the processes to the states that have fine structure levels differ not proportionally to the statistical populations of the states. 
The reason for that is the following. Transformation from the LS representation to the JJ representation changes parameters of each non-adiabatic region in the individual way.
Ultimately, this affects differently the rate coefficients.

It is seen from Figure~\ref{fig:neutr_rates} that the distribution of the mutual neutralization rate coefficients depends on the electron binding energy of the final channel. In turn, the dependence on the electron binding energy leads to the dependence on the excitation energy of the final channel. Finally, this indicates the existence of the optimal window. The channels 6 -- 13 belong to the optimal window and, hence, have the largest rate coefficients. Excitation and de-excitation processes due to transitions between the states from the optimal window have moderate values of the rate coefficients.
Nevertheless, the information about other partial processes is needed for NLTE modelling as well.

\begin{table}
\begin{center}
\renewcommand{\tabcolsep}{3.8pt}
\renewcommand{\arraystretch}{1.5}
\caption{SrH$^+$ (k~0$^+$) molecular states (in the $JJ$ representation), the corresponding scattering channels, and their asymptotic energies with respect to the ground-state level, as given by NIST \citep{NIST_ASD}.}
\label{tab:states} 
%\begin{small}
\begin{tabular}{llcc}
\hline
\hline
k &  \multicolumn{1}{c}{Scattering channels} & Asymptotic   \\ [-2mm] 
   &                                     & energies (eV)  \\[-1mm] 
\hline
1  & ${\rm Sr}^{+}\left(4p^{6}5s~{\rm^2S_{\rfrac{1}{2}}}\right) + {\rm H}(1s~{\rm^2S_{\rfrac{1}{2}}})$  &  0.0       \\[-1mm] \hline
2  & ${\rm Sr}^{+}\left(4p^{6}4d~{\rm^2D_{\rfrac{3}{2}}}\right) + {\rm H}(1s~{\rm^2S_{\rfrac{1}{2}}})$  &  1.8047016 \\ 
3  & ${\rm Sr}^{+}\left(4p^{6}4d~{\rm^2D_{\rfrac{5}{2}}}\right) + {\rm H}(1s~{\rm^2S_{\rfrac{1}{2}}})$  &  1.8394593 \\[-1mm] \hline
4  & ${\rm Sr}^{+}\left(4p^{6}5p~{\rm^2P^{\circ}_{\rfrac{1}{2}}}\right) + {\rm H}(1s~{\rm^2S_{\rfrac{1}{2}}})$  &  2.9403088 \\
5  & ${\rm Sr}^{+}\left(4p^{6}5p~{\rm^2P^{\circ}_{\rfrac{3}{2}}}\right) + {\rm H}(1s~{\rm^2S_{\rfrac{1}{2}}})$  &  3.0396772 \\[-1mm] \hline
6  & ${\rm Sr}^{+}\left(4p^{6}6s~{\rm^2S_{\rfrac{1}{2}}}\right) + {\rm H}(1s~{\rm^2S_{\rfrac{1}{2}}})$  &  5.9185754 \\[-1mm] \hline
7  & ${\rm Sr}^{+}\left(4p^{6}5d~{\rm^2D_{\rfrac{3}{2}}}\right) + {\rm H}(1s~{\rm^2S_{\rfrac{1}{2}}})$  &  6.6066604 \\ 
8  & ${\rm Sr}^{+}\left(4p^{6}5d~{\rm^2D_{\rfrac{5}{2}}}\right) + {\rm H}(1s~{\rm^2S_{\rfrac{1}{2}}})$  &  6.6174048 \\[-1mm] \hline
9  & ${\rm Sr}^{+}\left(4p^{6}6p~{\rm^2P^{\circ}_{\rfrac{1}{2}}}\right) + {\rm H}(1s~{\rm^2S_{\rfrac{1}{2}}})$  &  6.91456   \\ 
10 & ${\rm Sr}^{+}\left(4p^{6}6p~{\rm^2P^{\circ}_{\rfrac{3}{2}}}\right) + {\rm H}(1s~{\rm^2S_{\rfrac{1}{2}}})$  &  6.95029   \\[-1mm] \hline
11 & ${\rm Sr}^{+}\left(4p^{6}4f~{\rm^2F^{\circ}_{\rfrac{7}{2}}}\right) + {\rm H}(1s~{\rm^2S_{\rfrac{1}{2}}})$  &  7.561801  \\ 
12 & ${\rm Sr}^{+}\left(4p^{6}4f~{\rm^2F^{\circ}_{\rfrac{5}{2}}}\right) + {\rm H}(1s~{\rm^2S_{\rfrac{1}{2}}})$  &  7.561962  \\[-1mm] \hline
13 & ${\rm Sr}^{+}\left(4p^{6}7s~{\rm^2S_{\rfrac{1}{2}}}\right) + {\rm H}(1s~{\rm^2S_{\rfrac{1}{2}}})$  &  8.0545218 \\[-1mm] \hline
14 & ${\rm Sr}^{+}\left(4p^{6}6d~{\rm^2D_{\rfrac{3}{2}}}\right) + {\rm H}(1s~{\rm^2S_{\rfrac{1}{2}}})$  &  8.3717688 \\ 
15 & ${\rm Sr}^{+}\left(4p^{6}6d~{\rm^2D_{\rfrac{5}{2}}}\right) + {\rm H}(1s~{\rm^2S_{\rfrac{1}{2}}})$  &  8.3767629 \\[-1mm] \hline
16 & ${\rm Sr}^{+}\left(4p^{6}7p~{\rm^2P^{\circ}_{\rfrac{1}{2}}}\right) + {\rm H}(1s~{\rm^2S_{\rfrac{1}{2}}})$  &  8.515153  \\
17 & ${\rm Sr}^{+}\left(4p^{6}7p~{\rm^2P^{\circ}_{\rfrac{3}{2}}}\right) + {\rm H}(1s~{\rm^2S_{\rfrac{1}{2}}})$  &  8.532235  \\[-1mm] \hline
18 & ${\rm Sr}^{+}\left(4p^{6}5f~{\rm^2F^{\circ}_{\rfrac{5}{2}}}\right) + {\rm H}(1s~{\rm^2S_{\rfrac{1}{2}}})$  &  8.811036  \\ 
19 & ${\rm Sr}^{+}\left(4p^{6}5f~{\rm^2F^{\circ}_{\rfrac{7}{2}}}\right) + {\rm H}(1s~{\rm^2S_{\rfrac{1}{2}}})$  &  8.811036  \\[-1mm] \hline
20 & ${\rm Sr}^{+}\left(4p^{6}5g~{\rm^2G_{\rfrac{7}{2}}}\right) + {\rm H}(1s~{\rm^2S_{\rfrac{1}{2}}})$  &  8.847240  \\
21 & ${\rm Sr}^{+}\left(4p^{6}5g~{\rm^2G_{\rfrac{9}{2}}}\right) + {\rm H}(1s~{\rm^2S_{\rfrac{1}{2}}})$  &  8.847240  \\[-1mm] \hline
22 & ${\rm Sr}^{+}\left(4p^{6}8s~{\rm^2S_{\rfrac{1}{2}}}\right) + {\rm H}(1s~{\rm^2S_{\rfrac{1}{2}}})$  &  9.080243  \\[-1mm] \hline
23 & ${\rm Sr}^{+}\left(4p^{6}7d~{\rm^2D_{\rfrac{3}{2}}}\right) + {\rm H}(1s~{\rm^2S_{\rfrac{1}{2}}})$  &  9.251862  \\ 
24 & ${\rm Sr}^{+}\left(4p^{6}7d~{\rm^2D_{\rfrac{5}{2}}}\right) + {\rm H}(1s~{\rm^2S_{\rfrac{1}{2}}})$  &  9.254565  \\[-1mm] \hline
25 & ${\rm Sr}^{+}\left(4p^{6}8p~{\rm^2P^{\circ}_{\rfrac{3}{2}}}\right) + {\rm H}(1s~{\rm^2S_{\rfrac{1}{2}}})$  &  9.337473  \\[-1mm] \hline
26 & ${\rm Sr}^{+}\left(4p^{6}6f~{\rm^2F^{\circ}_{\rfrac{7}{2}}}\right) + {\rm H}(1s~{\rm^2S_{\rfrac{1}{2}}})$  &  9.491412  \\ 
27 & ${\rm Sr}^{+}\left(4p^{6}6f~{\rm^2F^{\circ}_{\rfrac{5}{2}}}\right) + {\rm H}(1s~{\rm^2S_{\rfrac{1}{2}}})$  &  9.491412  \\[-1mm] \hline
28 & ${\rm Sr}^{+}\left(4p^{6}6g~{\rm^2G_{\rfrac{7}{2}}}\right) + {\rm H}(1s~{\rm^2S_{\rfrac{1}{2}}})$  &  9.514262  \\ 
29 & ${\rm Sr}^{+}\left(4p^{6}6g~{\rm^2G_{\rfrac{9}{2}}}\right) + {\rm H}(1s~{\rm^2S_{\rfrac{1}{2}}})$  &  9.514262  \\[-1mm] \hline
30 & ${\rm Sr}^{+}\left(4p^{6}9s~{\rm^2S_{\rfrac{1}{2}}}\right) + {\rm H}(1s~{\rm^2S_{\rfrac{1}{2}}})$  &  9.653112  \\[-1mm] \hline
31 & ${\rm Sr}^{2+}\left(4p^{6}~{\rm^1S_{0}}\right) + {\rm H}^-(1s^2~{\rm^1S_{0}})$     & 10.2762764   \\[-1mm]
\hline 
\hline
\end{tabular}
\end{center}
\end{table}

\subsection{Codes and model ingredients}\label{Sect:codes}

The coupled radiative transfer and statistical equilibrium equations were solved with   the code {\sc detail} \citep{detail}, where the opacity package was revised as presented by
\citet{mash_fe}. The departure coefficients, $b_i = n_i^{\rm NLTE}/n_i^{\rm LTE}$, yielded by {\sc detail} are then used by the code {\sc synthV}\_NLTE \citep{2019ASPC} to calculate the NLTE line profiles for a given NLTE species. Here, $n_i^{\rm NLTE}$ and $n_i^{\rm LTE}$ are the statistical equilibrium and thermal (Saha-Boltzmann) number densities, respectively. Spectral lines of the other species are computed under the LTE assumption. The line list required for calculations of the synthetic spectra, together with atomic data, were taken from the Vienna Atomic Line Database \citep[VALD,][]{2015PhyS...90e4005R}. The best fit to the observed spectrum is obtained using the {\sc IDL binmag} code by O.~Kochukhov\footnote{http://www.astro.uu.se/$\sim$oleg/binmag.html}.

We adopt the same list of spectral lines and their atomic parameters, namely, $gf$-values, van der Waals broadening constants, isotopic splitting (IS), and hyper-fine splitting (HFS) structure, as in previous papers of the series \citep{dsph_parameters,2017A&A...608A..89M,2019AstL...45..259P,2021_coma}. The spectral lines used in the abundance analysis of individual stars are listed in Table\,\ref{Tab:linelist}. 
  
As previously, we used the MARCS model structures \citep{Gustafssonetal:2008}.

\begin{table} % [htbp]
 \centering
 \caption{\label{Tab:linelist} Line atomic data, equivalent width, LTE, and NLTE abundances, $\eps{}$, for individual lines in the sample stars.} 
 \begin{tabular}{lccrrcc}
\hline\hline \noalign{\smallskip}
Species & $\lambda$ & \Eexc & log~$gf$ & \multicolumn{1}{c}{EW$^1$} & LTE & NLTE \\
        & [\AA]     & [eV]  &       & \multicolumn{1}{c}{[m\AA]} &     &      \\
\noalign{\smallskip} \hline \noalign{\smallskip}
\multicolumn{7}{l}{S04-130} \\
\ion{Na}{i} &  5889.95 & 0.00 &  0.12 & 186.0 & 3.64 & 3.20  \\
\ion{Na}{i} &  5895.92 & 0.00 & -0.19 & 157.8 & 3.53 & 3.14  \\
\ion{Mg}{i} &  5172.68 & 2.71 & -0.45 & 204.5 & 5.06 & 5.06  \\
\ion{Mg}{i} &  5528.40 & 4.35 & -0.50 &  60.6 & 4.99 & 4.98 \\
\noalign{\smallskip}\hline \noalign{\smallskip}
\multicolumn{7}{l}{{\bf Notes.} $^1$ Equivalent width corresponds to the best } \\
\multicolumn{7}{l}{theoretical fit. This table is available in its entirety in} \\
\multicolumn{7}{l}{a machine-readable form in the online journal. A portion is} \\
%\multicolumn{7}{l}{} \\
\multicolumn{7}{l}{shown here for guidance regarding its form and content.} \\
\noalign{\smallskip} \hline
\end{tabular}
\end{table}

\section{Abundance results}\label{sect:abundances}

A number of chemical elements, for which their NLTE and/or LTE abundances were derived in this study, ranges between 13 and 7 depending on the spectral coverage and S/N of the observed spectra available. The abundances obtained from individual lines in each individual star are presented in Table~\ref{Tab:linelist}. The microturbulent velocities ($\xi_t$) needed for abundance determinations were calculated using the empirical formula obtained in Paper~I for metal-poor Galactic halo giants. It provides an accuracy of about 0.2~\kms. 
Table~\ref{Tab:abund} lists the elemental average LTE and NLTE abundances together with their 
%statistical errors and the number of lines used, $N_l$. 
%The statistical error is meant as the 
dispersions in the single line measurements around the mean, $\sigma = \sqrt{\Sigma(\overline{x}-x_i)^2 / (N_l-1)}$, and the number of lines used, $N_l$. For the species with a single line used in abundance analysis, the stochastic error caused by random uncertainties in the continuum placement and line profile fitting is estimated as $\sigma$ = 0.1~dex.
For consistency with our previous studies, we used the solar photosphere abundances, $\eps{\odot}$, from \citet{AG1989}. The exceptions are Ti and Fe, for which we adopted $\eps{Ti,\odot}$ = 4.93 \citep[][meteoritic abundance]{Lodders2009} and $\eps{Fe,\odot}$ = 7.50 \citep{1998SSRv...85..161G}. 
In order to compute the metallicities, [Fe/H], presented in Table~\ref{Tab:parameters}, and the abundance ratios relative to iron, [X/Fe], we averaged the NLTE abundances derived from the Fe\ione\ and Fe\ii\ lines. 

\begin{table*} % [htbp]
 \centering
 \caption{\label{Tab:abund} Summary of the LTE and NLTE abundances of the investigated stars.} 
 \begin{tabular}{clcrrrcr|rrrcr}
\hline\hline \noalign{\smallskip}
 Z & Atom & $\eps{\odot}$ & $N_l$ & \multicolumn{1}{c}{LTE} & \multicolumn{3}{c|}{NLTE} & $N_l$ & \multicolumn{1}{c}{LTE} & \multicolumn{3}{c}{NLTE} \\
\cline{6-8} 
\cline{11-13} \noalign{\smallskip}
   &      &               &       & \multicolumn{1}{c}{$\eps{}$} & \multicolumn{1}{c}{$\eps{}$} & [X/H] & [X/Fe] &       & \multicolumn{1}{c}{$\eps{}$} & \multicolumn{1}{c}{$\eps{}$} & [X/H] & [X/Fe] \\
\noalign{\smallskip} \hline \noalign{\smallskip}
    &      &               & \multicolumn{5}{l}{ \ \ S~04-130} &  \multicolumn{5}{l}{ \ \ S~11-97} \\
  6 & CH    &  8.39 &   3 &  5.14(0.14) &             & -3.25 & -0.60  &   2 &  5.14(0.10) &             & -3.25 & -0.63 \\ 
 11 & Na I  &  6.33 &   2 &  3.59(0.08) &  3.17(0.04) & -3.16 & -0.44 &   2 &  3.79(0.02) &  3.31(0.05) & -3.02 & -0.32 \\
 12 & Mg I  &  7.58 &   3 &  5.04(0.04) &  5.03(0.04) & -2.55 &  0.17 &   3 &  5.15(0.06) &  5.15(0.06) & -2.43 &  0.27 \\
 13 & Al I  &  6.47 &   1 &  2.98(0.10) &  3.04(0.10) & -3.43 & -0.71 &   1 &  3.08(0.10) &  3.13(0.10) & -3.34 & -0.64 \\
 14 & Si I  &  7.55 &   2 &  5.05(0.11) &  5.04(0.11) & -2.51 &  0.21 &   1 &  5.07(0.10) &  5.06(0.10) & -2.49 &  0.21 \\
 20 & Ca I  &  6.36 &   6 &  3.70(0.12) &  3.84(0.09) & -2.52 &  0.20 &   7 &  3.74(0.12) &  3.88(0.07) & -2.48 &  0.22 \\
 22 & Ti I  &  4.93 &  10 &  2.10(0.07) &  2.39(0.07) & -2.54 &  0.18 &  10 &  2.21(0.11) &  2.46(0.10) & -2.47 &  0.23 \\
 22 & Ti II &  4.93 &  19 &  2.26(0.14) &  2.27(0.15) & -2.66 &  0.06 &  18 &  2.33(0.13) &  2.34(0.14) & -2.59 &  0.11 \\
 26 & Fe I  &  7.50 &  36 &  4.60(0.14) &  4.74(0.15) & -2.76 & -0.04 &  36 &  4.63(0.16) &  4.77(0.15) & -2.73 & -0.03 \\
 26 & Fe II &  7.50 &   5 &  4.81(0.09) &  4.81(0.09) & -2.69 &  0.03 &   5 &  4.83(0.05) &  4.83(0.05) & -2.67 &  0.03 \\
 28 & Ni I  &  6.25 &   4 &  3.39(0.26) &             &       &  0.04 &   3 &  3.51(0.05) &             &       &  0.13 \\
 30 & Zn I  &  4.60 &   1 &  2.25(0.10) &  2.14(0.10) & -2.46 &  0.26 &   1 &  2.17(0.10) &  2.06(0.10) & -2.54 &  0.16 \\
 38 & Sr II &  2.90 &   1 & -0.56(0.10) & -0.51(0.10) & -3.41 & -0.69 &   1 & -0.87(0.10) & -0.83(0.10) & -3.73 & -1.03 \\
 40 & Zr II &  2.56 &   1 & -0.36(0.10) & -0.24(0.10) & -2.80 & -0.08 &   1 & -0.02(0.10) &  0.10(0.10) & -2.46 &  0.24 \\
 56 & Ba II &  2.13 &   2 & -1.54(0.02) & -1.50(0.01) & -3.63 & -0.91 &   2 & -1.45(0.01) & -1.41(0.03) & -3.54 & -0.84 \\
    &      &               & \multicolumn{5}{l}{ \ \ S~11-04} &  \multicolumn{5}{l}{ \ \ S~24-72} \\
  6 & CH    & 8.39 &   1 &  4.79(0.20) &             & -3.60 & -0.97 &   1 &  6.12(0.20)  &             & -2.27 &  0.55 \\ 
 11 & Na I  & 6.33 &   2 &  3.48(0.02) &  3.33(0.03) & -3.00 & -0.40 &   2 &  4.48(0.11)  &  4.34(0.06) & -1.99 &  0.85 \\
 12 & Mg I  & 7.58 &   3 &  5.17(0.02) &  5.22(0.07) & -2.36 &  0.24 &   3 &  4.86(0.05)  &  4.91(0.07) & -2.67 &  0.17 \\
 13 & Al I  & 6.47 &   1 &  3.12(0.20) &  3.32(0.20) & -3.15 & -0.55 &   1 &  2.76(0.20)  &  2.96(0.20) & -3.51 & -0.67 \\
 14 & Si I  & 7.55 &   1 &  5.16(0.20) &  5.16(0.20) & -2.39 &  0.21 &   1 &  4.69(0.20)  &  4.69(0.20) & -2.86 & -0.02 \\
 20 & Ca I  & 6.36 &   6 &  3.75(0.16) &  3.90(0.16) & -2.46 &  0.14 &   4 &  3.59(0.10)  &  3.77(0.14) & -2.59 &  0.25 \\
 22 & Ti I  & 4.93 &   7 &  2.02(0.06) &  2.49(0.13) & -2.44 &  0.16 &   6 &  1.85(0.08)  &  2.22(0.10) & -2.71 &  0.13 \\
 22 & Ti II & 4.93 &   7 &  2.40(0.10) &  2.42(0.10) & -2.51 &  0.09 &   7 &  2.23(0.10)  &  2.24(0.10) & -2.69 &  0.15 \\
 26 & Fe I  & 7.50 &  37 &  4.68(0.14) &  4.84(0.13) & -2.66 & -0.06 &  43 &  4.55(0.14)  &  4.71(0.14) & -2.79 &  0.05 \\
 26 & Fe II & 7.50 &   4 &  4.90(0.12) &  4.90(0.12) & -2.60 &  0.00 &   3 &  4.66(0.05)  &  4.66(0.05) & -2.84 &  0.00 \\
 28 & Ni I  & 6.25 &   1 &  3.22(0.10) &             &       & -0.21 &   1 &  3.33(0.10)  &             &       &  0.03 \\
 38 & Sr II & 2.90 &   2 & -0.39(0.15) & -0.20(0.14) & -3.10 & -0.50 &   1 & -0.54(0.10)  & -0.36(0.20) & -3.26 & -0.42 \\
 56 & Ba II & 2.13 &   2 & -1.59(0.00) & -1.49(0.05) & -3.62 & -1.02 &   1 & -1.92(0.10)  & -1.79(0.20) & -3.92 & -1.08 \\
    &      &               & \multicolumn{5}{l}{ \ \ S~10-14} &  \multicolumn{5}{l}{ \ \ S~11-13} \\
 11 & Na I  &  6.33 &   1 &  2.85(0.10) &  2.70(0.16) & -3.63 & -0.66  &   2 &  3.03(0.13) &  2.77(0.17) & -3.56 & -0.51 \\
 12 & Mg I  &  7.58 &   1 &  4.90(0.10) &  4.96(0.10) & -2.62 &  0.35  &   2 &  4.86(0.03) &  4.88(0.08) & -2.70 &  0.35 \\
 20 & Ca I  &  6.36 &   4 &  3.53(0.13) &  3.67(0.13) & -2.69 &  0.28  &   5 &  3.49(0.14) &  3.63(0.14) & -2.73 &  0.32 \\
 22 & Ti I  &  4.93 &   4 &  1.85(0.12) &  2.15(0.13) & -2.78 &  0.19  &   3 &  1.91(0.07) &  2.21(0.07) & -2.72 &  0.33 \\
 22 & Ti II &  4.93 &   3 &  1.95(0.13) &  1.97(0.14) & -2.96 &  0.01  &   3 &  2.04(0.09) &  2.06(0.10) & -2.87 &  0.18 \\
 26 & Fe I  &  7.50 &  28 &  4.43(0.16) &  4.62(0.15) & -2.88 &  0.09  &  25 &  4.36(0.11) &  4.53(0.11) & -2.97 &  0.08 \\
 26 & Fe II &  7.50 &   8 &  4.44(0.16) &  4.44(0.16) & -3.06 & -0.09  &   7 &  4.37(0.14) &  4.37(0.14) & -3.13 & -0.08 \\
 28 & Ni I  &  6.25 &   1 &  3.00(0.10) &             & -3.25 & -0.18  &   1 &  2.97(0.10) &             & -3.28 & -0.14 \\
 56 & Ba II &  2.13 &   2 & -1.88(0.04) & -1.84(0.01) & -3.97 & -1.00  &   2 & -1.68(0.03) & -1.60(0.01) & -3.73 & -0.68 \\
    &      &               & \multicolumn{5}{l}{ \ \ S~11-37} & \multicolumn{5}{l}{ \ \ S~12-28} \\
 11 & Na I  &  6.33 &   2 &  3.19(0.08) &  2.95(0.15) & -3.38 & -0.49  &   2 &  3.34(0.19) &  3.10(0.13) & -3.23 & -0.37 \\
 12 & Mg I  &  7.58 &   4 &  4.76(0.21) &  4.78(0.16) & -2.80 &  0.09  &   5 &  5.19(0.08) &  5.28(0.11) & -2.30 &  0.55 \\
 20 & Ca I  &  6.36 &   3 &  3.87(0.09) &  3.97(0.10) & -2.39 &  0.50  &   4 &  3.85(0.22) &  3.95(0.17) & -2.41 &  0.44 \\
 22 & Ti I  &  4.93 &   2 &  2.07(0.03) &  2.34(0.03) & -2.59 &  0.30  &   5 &  2.03(0.13) &  2.32(0.14) & -2.61 &  0.24 \\
 22 & Ti II &  4.93 &   4 &  2.15(0.23) &  2.17(0.23) & -2.76 &  0.13  &   4 &  2.21(0.13) &  2.22(0.12) & -2.71 &  0.15 \\
 26 & Fe I  &  7.50 &  20 &  4.45(0.14) &  4.63(0.14) & -2.87 &  0.02  &  22 &  4.49(0.15) &  4.66(0.15) & -2.84 &  0.01 \\
 26 & Fe II &  7.50 &   2 &  4.59(0.05) &  4.59(0.05) & -2.91 & -0.02  &   4 &  4.63(0.10) &  4.63(0.10) & -2.87 & -0.01 \\
 28 & Ni I  &  6.25 &   1 &  3.43(0.10) &             & -2.82 &  0.23  &   1 &  2.96(0.10) &             & -3.29 & -0.28 \\
 56 & Ba II &  2.13 &   2 & -1.90(0.10) & -1.86(0.15) & -3.99 & -1.10  &   1 & -1.17(0.10) & -1.12(0.10) & -3.25 & -0.39 \\
    &      &               & \multicolumn{5}{l}{ \ \ S~14-98} &  \multicolumn{5}{l}{ \ \ S~15-19} \\
  6 & CH    &  8.39 &     &             &             &       &        &   4 &  5.39(0.25) &             & -3.00 &  0.28 \\ 
 11 & Na I  &  6.33 &   1 &  2.96(0.00) &  2.82(0.10) & -3.51 & -0.78  &   2 &  3.14(0.07) &  2.95(0.00) & -3.38 & -0.06 \\
 12 & Mg I  &  7.58 &   1 &  4.93(0.10) &  4.91(0.10) & -2.67 &  0.06  &   3 &  4.80(0.18) &  4.78(0.13) & -2.80 &  0.52 \\
 20 & Ca I  &  6.36 &   3 &  3.81(0.17) &  3.91(0.18) & -2.45 &  0.28  &   6 &  3.59(0.19) &  3.75(0.15) & -2.61 &  0.71 \\
 22 & Ti I  &  4.93 &   4 &  2.34(0.14) &  2.60(0.15) & -2.33 &  0.40  &   2 &  1.76(0.09) &  2.11(0.08) & -2.82 &  0.50 \\
 22 & Ti II &  4.93 &   3 &  2.53(0.15) &  2.54(0.16) & -2.39 &  0.34  &   4 &  2.03(0.27) &  2.07(0.28) & -2.86 &  0.46 \\
 26 & Fe I  &  7.50 &  13 &  4.63(0.14) &  4.76(0.14) & -2.74 & -0.01  &  13 &  4.02(0.08) &  4.23(0.07) & -3.27 &  0.05 \\
 26 & Fe II &  7.50 &   2 &  4.78(0.00) &  4.78(0.10) & -2.72 &  0.01  &   4 &  4.12(0.14) &  4.12(0.14) & -3.38 & -0.06 \\
 28 & Ni I  &  6.25 &   1 &  3.24(0.10) &             & -3.01 & -0.14  &   1 &  2.69(0.10) &             & -3.56 & -0.08 \\
 56 & Ba II &  2.13 &   2 & -1.70(0.10) & -1.66(0.15) & -3.79 & -1.06  &   2 & -0.57(0.10) & -0.58(0.15) & -2.71 &  0.61 \\
\hline \noalign{\smallskip}
\end{tabular}
\end{table*}
 
\begin{table*} % [htbp]
 \centering
 \contcaption{\label{tab:continued} -- Summary of the LTE and NLTE abundances of the investigated stars.} 
 \begin{tabular}{clcrrrcr|rrrcr}
\hline\hline \noalign{\smallskip}
 Z & Atom & $\eps{\odot}$ & $N_l$ & \multicolumn{1}{c}{LTE} & \multicolumn{3}{c|}{NLTE} & $N_l$ & \multicolumn{1}{c}{LTE} & \multicolumn{3}{c}{NLTE} \\
\cline{6-8} 
\cline{11-13} \noalign{\smallskip}
   &      &               &       & \multicolumn{1}{c}{$\eps{}$} & \multicolumn{1}{c}{$\eps{}$} & [X/H] & [X/Fe] &       & \multicolumn{1}{c}{$\eps{}$} & \multicolumn{1}{c}{$\eps{}$} & [X/H] & [X/Fe] \\
\noalign{\smallskip} \hline \noalign{\smallskip}
    &      &               & \multicolumn{5}{l}{ \ \ S~49} & \multicolumn{5}{l}{ } \\
 12 & Mg I  &  7.58 &   3 &  4.92(0.14) &  4.98(0.12) & -2.60 &  0.42 &  &  &  &  &  \\
 20 & Ca I  &  6.36 &   1 &  3.56(0.10) &  3.85(0.10) & -2.51 &  0.51 &  &  &  &  &  \\
 22 & Ti I  &  4.93 &   5 &  2.01(0.10) &  2.33(0.09) & -2.60 &  0.42 &  &  &  &  &  \\
 22 & Ti II &  4.93 &   4 &  2.18(0.14) &  2.20(0.15) & -2.73 &  0.29 &  &  &  &  &  \\
 26 & Fe I  &  7.50 &  14 &  4.34(0.14) &  4.50(0.14) & -3.00 &  0.02 &  &  &  &  &  \\
 26 & Fe II &  7.50 &   5 &  4.46(0.09) &  4.46(0.09) & -3.04 & -0.02 &  &  &  &  &  \\
 28 & Ni I  &  6.25 &   1 &  2.85(0.10) &             & -3.40 & -0.24 &  &  &  &  &  \\
 56 & Ba II &  2.13 &   1 & -1.74(0.10) & -1.66(0.10) & -3.79 & -0.77 &  &  &  &  &  \\
\hline \noalign{\smallskip}
\multicolumn{13}{l}{{\bf Notes.} The numbers in parentheses are the dispersions in the single line measurements around the mean. For $N_l$ = 1, } \\
\multicolumn{13}{l}{$\sigma$ = 0.1~dex was adopted. [Ni/Fe] was computed using the LTE abundances from lines of Ni\ione\ and Fe\ione.} \\
\noalign{\smallskip} \hline
\end{tabular}
\end{table*}

Below we describe abundance determinations for individual elements.

\subsection{Metallicities}\label{Sect:feh}

Following Paper~I, we did not use 
%We used the same set of Fe\ione\ and Fe\ii\ lines together with their $gf$-values and van der Waals broadening constants, as in Paper~I. It does not include 
the Fe\ione\ lines with low excitation energy of the lower level, \Eexc\ $<$ 1.2~eV, that are predicted to be affected by hydrodynamic phenomena (3D effects) in the atmosphere to more degree than the higher excitation lines \citep[for example,][]{Collet2007,2013A&A...559A.102D}. In a given star, we did not use the lines with EW $>$ 120~m\AA, in order to minimize the sphericity effects on derived abundances. The fact is that we employ the MARCS spherical atmosphere models \citep{Gustafssonetal:2008}, 
%as provided by the MARCS web site\footnote{\tt   http://marcs.astro.uu.se}, 
while all our codes treat the radiation transfer in plane-parallel geometry. As discussed in Paper~I, such an inconsistent approach results in abundance shift, which depends only on the line strength and  does not exceed 0.06~dex for the EW $<$ 120~m\AA\ lines. 

\begin{figure}  %[htbp]
 \begin{minipage}{85mm}
\centering
	\includegraphics[width=0.99\textwidth, clip]{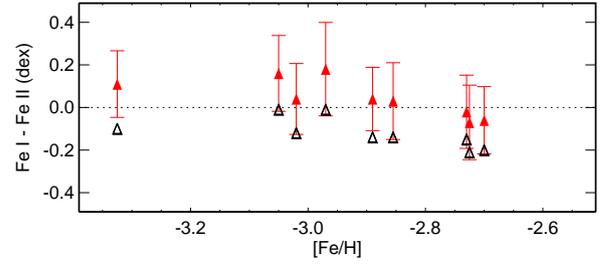}
  \caption{\label{Fig:fe12} Abundance differences between the two ionisation stages for iron, Fe\ione\ - Fe\ii\ = $\eps{FeI} - \eps{FeII}$, for the Sextans dSph stars. The filled and open triangles correspond to the NLTE and LTE calculations, respectively. The error bars correspond to $\sigma_{\rm FeI - FeII} = \sqrt{\sigma_{\rm FeI}^2 + \sigma_{\rm FeII}^2}$ for NLTE.}
\end{minipage}
\end{figure}

%Sextans, Fe I- Fe II (NLTE) =    0.0411111+-    0.0963645(       9 stars)
%Sextans, Fe I- Fe II (LTE) =    -0.125556+-    0.0798610(       9 stars)

It was found that the LTE abundances from lines of Fe\ione\ are systematically lower compared with that for Fe\ii, see Fig.~\ref{Fig:fe12}. NLTE leads to weakened Fe\ione\ lines and positive NLTE abundance corrections, resulting in consistent abundances from the two ionisation stages, with $\eps{FeI} - \eps{FeII}$ = 0.05$\pm$0.09. 
%Statistical error of the abundance difference computed as $\sigma_{\rm FeI - FeII} = \sqrt{\sigma_{\rm FeI}^2 + \sigma_{\rm FeII}^2}$ ranges between 0.15~dex and 0.22~dex for different stars. 

Thus, the effective temperatures and surface gravities derived in this study from photometric methods are supported spectroscopically. 

\subsection{Carbon}

The C abundances are reported in the literature for four stars of our sample. Two of them, S~04-130 and S~11-97, are carbon-normal cool giants (LLP20). Two another stars, S~24-72 and S~15-19, are moderately enhanced in C, with [C/Fe] = 0.40 \citep{2010A&A...524A..58T} and [C/Fe] = 1.0 \citep{2011PASJ...63S.523H}. Our results based on analysis of the molecular CH 4310-4313, 4322.5-4324.5, and 4366.0-4367.0~\AA\ bands support the earlier determinations, although we obtained lower [C/Fe] = 0.28 for S~15-19. 
%was derived for the three stars, using  The stars S~04-130 and S~11-97, with their [C/Fe] = $-0.60$ and $-0.63$, respectively, appear carbon-normal cool giants, according to a definition of \citet{2007ApJ...655..492A}. A carbon enhancement of [C/Fe] = 0.28 was obtained for S~15-19. Applying the carbon correction from \citet{2014ApJ...797...21P} leads to an initial [C/Fe] $\simeq$ 1. 
An origin of carbon enhancement is, probably, different for S~24-72 and S~15-19. The star S~24-72 is enhanced in C and Na, but not in the neutron-capture (n-capture) elements, while S~15-19, in contrast, has the Na abundance, which is typical for 'normal' VMP stars, and is enhanced in Ba (see below).

\subsection{$\alpha$-process elements}

\begin{figure*}
 \begin{minipage}{170mm}
\centering
\hspace{-15mm}
\parbox{0.23\linewidth}{\includegraphics[scale=0.35]{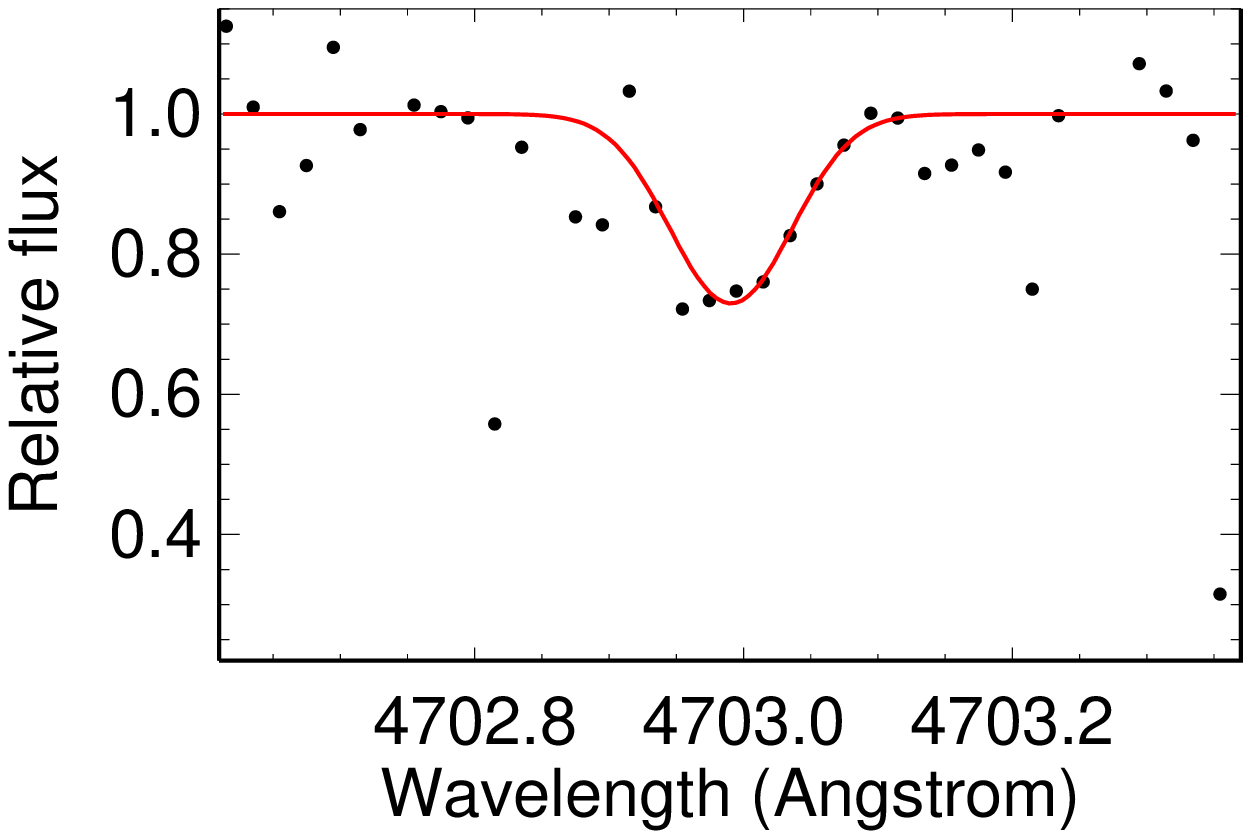}\\
\centering}
\hspace{3mm}
\parbox{0.23\linewidth}{\includegraphics[scale=0.35]{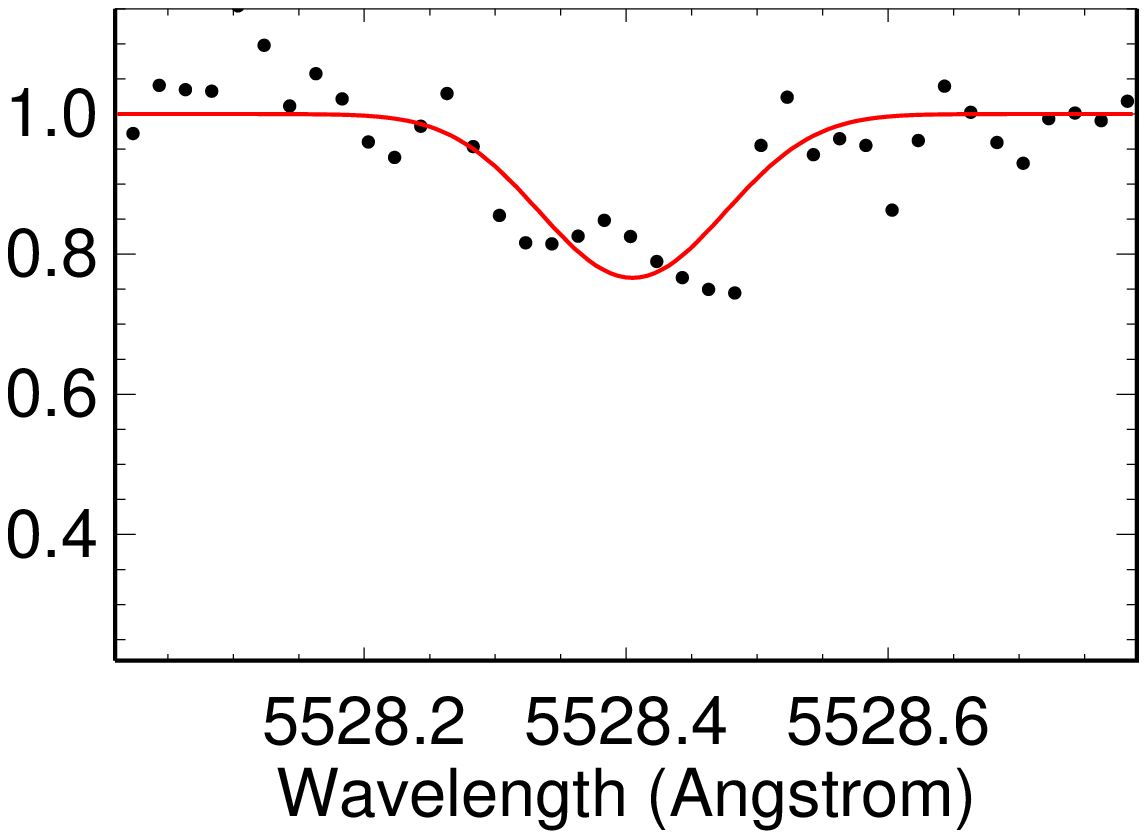}\\
\centering}
\hspace{3mm}
\parbox{0.23\linewidth}{\includegraphics[scale=0.35]{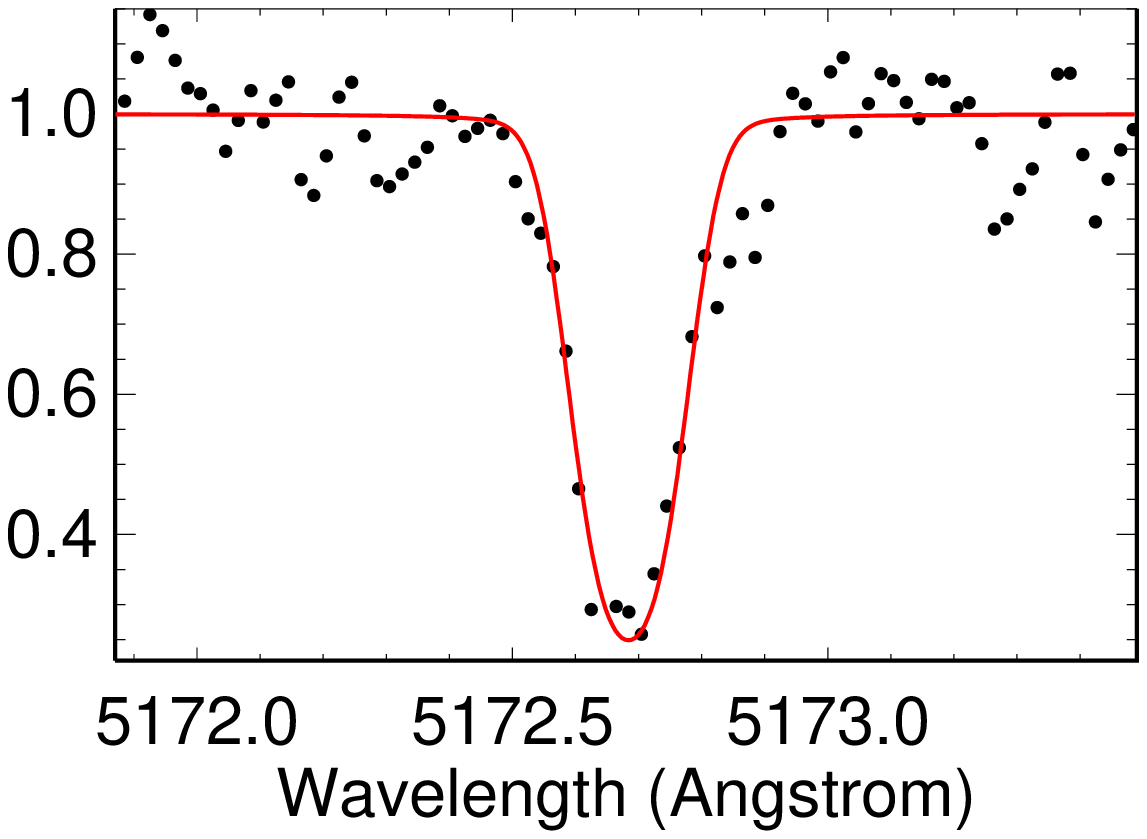}\\
\centering}
\hspace{3mm}
\parbox{0.23\linewidth}{\includegraphics[scale=0.35]{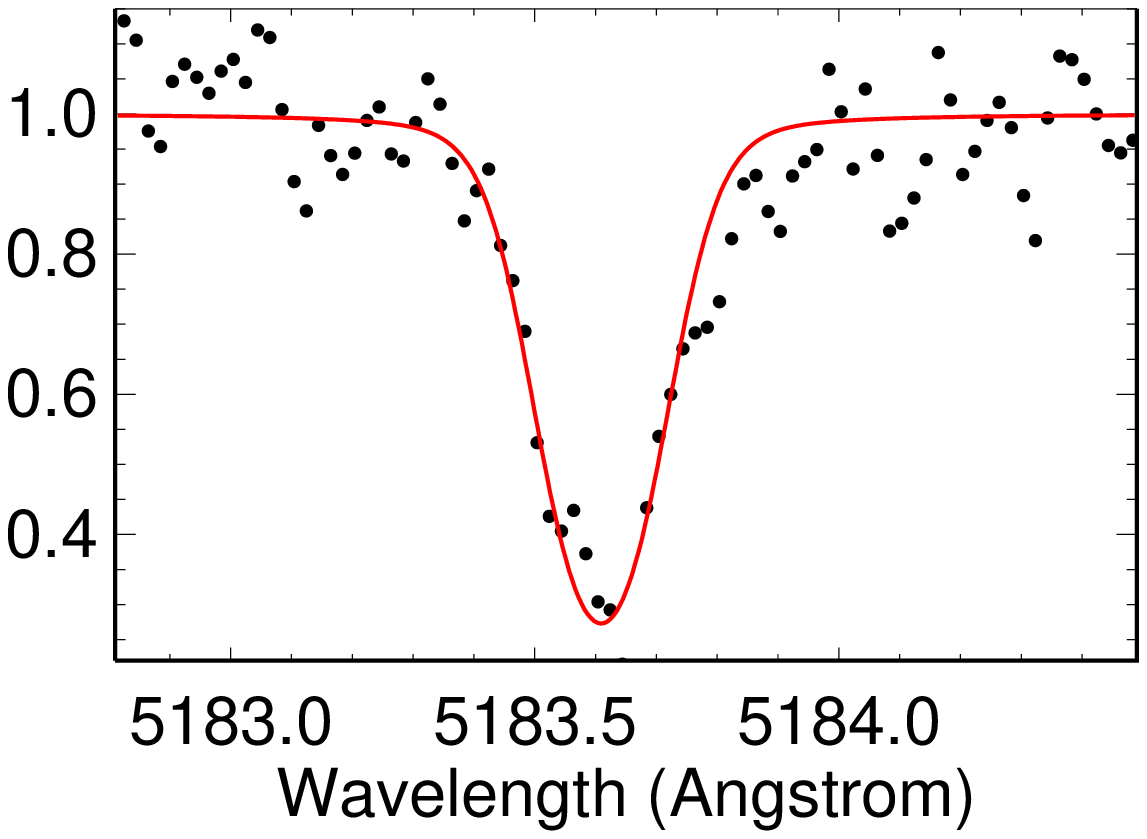}\\
\centering}
\vspace{-3mm}
    \caption{Best NLTE fits (continuous curves) to the observed Mg\ione\ 4703, 5528, 5172, and 5183\,\AA\ lines (panels from left to right) in S~11-37 (bold dots). The obtained NLTE abundances are $\eps{}$ = 4.75, 4.84, 4.53, and 4.92, respectively.}
    \label{fig:mg_lines}
\end{minipage}
\end{figure*}

Magnesium, calcium, and titanium were measured in each star of our sample. The best observed among them is Ca, with the three to seven Ca\ione\ lines of moderate strength in different stars. The exception is S~49, with the only Ca\ione\ 4425~\AA\ measured. For Mg, we used all the lines, which can be measured, including strong (EW $>$ 120~m\AA) Mg\ione b lines. For a given star with more than one Mg\ione\ line, the NLTE abundances from different lines were obtained to be consistent. See, for example, Fig.~\ref{fig:mg_lines} for S~11-37. 

Titanium was observed in lines of Ti\ione\ and Ti\ii. For the majority of our sample stars, the average LTE abundance from lines of Ti\ione\ appears to be lower than that for Ti\ii. NLTE leads to weakened lines of Ti\ione\ and positive NLTE abundance corrections of $\Delta_{\rm NLTE} = \eps{\rm NLTE} - \eps{\rm LTE}$ = 0.20~dex to 0.35~dex, depending on atmospheric parameters. While the NLTE effects are minor for Ti\ii, with slightly negative or slightly positive $\Delta_{\rm NLTE}$. For S~04-130 and S~11-97, with 10 lines of Ti\ione\ and 18 lines of Ti\ii\ in each star, NLTE leads to consistent abundances from lines of the two ionization stages, within 0.05~dex and 0.08~dex, respectively. In the Subaru spectra of the remaining stars, we could measure two to five lines of Ti\ione\ and three to four lines of Ti\ii. For these stars, the difference in NLTE abundances between Ti\ione\ and Ti\ii\ amounts to 0.06 to 0.28~dex. When inspecting the [Ti/Fe] abundance trend in the Sextans dSph (Sect.~\ref{sect:trends}), we rely on the Ti\ii-based NLTE abundances.

\subsection{Neutron-capture elements}\label{sect:heavy}

Only for barium we could derive its abundance in each star of our sample. Both Ba\ii\ 4934\,\AA\ resonance line and Ba\ii\ 5853, 6141, and 6497\,\AA\ subordinate lines were measured, where available. The Ba\ii\ 4554\,\AA\ line is covered by the HDS/Subaru spectra, however, is heavily affected by noise and was not used in abundance analysis. The abundance derived from the Ba\ii\ resonance lines depends on the Ba isotope mixture adopted in the calculations because of substantial separations between the HFS components. The odd atomic mass isotopes $^{135}$Ba and $^{137}$Ba have very similar HFS \citep{Brix1952,1980ZPhyA.295..311S,1982PhRvA..25.1476B, 1983ZPhyA.311...41B}. Therefore, the abundance is essentially dependent on the total fractional abundance of these odd isotopes, $f_{\rm odd}$. In the solar system matter, $f_{\rm odd}$ = 0.18 \citep{Lodders2009}, while it is larger for pure r-process production of barium and varies between $f_{\rm odd}$ = 0.438 \citep{Kratz2007} and 0.72 \citep{McWilliam1998}. For example, for Ba\ii\ 4934\,\AA\ in S~04-130, moving from $f_{\rm odd}$ = 0.18 to $f_{\rm odd}$ = 0.44 leads to a 0.15~dex lower Ba abundance. The Ba\ii\ subordinate lines are almost free of HFS effects, and they were used to obtain the final Ba abundances for Table~\ref{Tab:abund}. The exception is S~49, with only Ba\ii\ 4934\,\AA\ available. In this latter case, we applied $f_{\rm odd}$ = 0.46, which corresponds to the r-process production of Ba, as predicted by \citet{Arlandini1999}.

Stellar $f_{\rm odd}$ value is an indicator of whether r- or s-process dominated the Ba production at the time when the star formed. In the absence of the Eu abundances, determination of $f_{\rm odd}$ is  important for understanding synthesis of heavy elements at the epochs, when the Sextans galaxy was very metal-poor.
We determined abundances from Ba\ii\ 4934\,\AA\ using $f_{\rm odd}$ = 0.18 and $f_{\rm odd}$ = 0.46. Together with S~11-04 and S~24-72 from Paper~II, we have nine Sextans stars with both the resonance and subordinate lines available. In case of $f_{\rm odd}$ = 0.46, abundances from the resonance and subordinate lines are consistent within 0.08~dex for the two stars only, while the remaining stars, including the four stars with the UVES/VLT spectra available and $\sigma_{\rm Ba} \le$ 0.05~dex, reveal negative abundance difference $\Delta(res - sub) = \eps{res} - \eps{sub} \le -0.16$. In case of the solar mixture of the Ba isotopes, $\Delta(res - sub)$ exceeds 0.10~dex (NLTE) for the only star, S~14-98, and this abundance difference is at the level of 1$\sigma_{\rm Ba}$. These abundance comparisons between Ba\ii\ 4934\,\AA\ and the subordinate lines favour the s-process as 
%for the six stars. We note that an important source of the uncertainty in the abundance derived from Ba\ii\ 4934\,\AA\ is microturbulent velocity. In our sample stars, this line has EW $\sim$ 100~m\AA, and a variation of 0.2~\kms\ in $\xi_t$ changes the abundance by 0.07~dex. Thus, a difference of 0.11~dex is within the error bars of abundance determinations. These abundance comparisons can be considered as a hint of the s-process being 
a major contributor to Ba production in the Sextans dSph because, in the solar barium with $f_{\rm odd}$ = 0.18, a fraction of s-nuclei amounts to 81\%\ \citep{Arlandini1999}. Interestingly, an evidence for dominant contribution of the s-process to an early enrichment in Ba was also provided for the Sculptor dSph \citep{2015A&A...583A..67J}.

Our sample stars reveal low abundances of Ba, with [Ba/Fe] $\sim -1$, in line with the earlier findings for the VMP stars in dwarf galaxies \citep{2001ApJ...548..592S,2009A&A...502..569A,2010ApJ...719..931C,2012AJ....144..168K,2015A&A...583A..67J,2017A&A...608A..89M,2020A&A...636A.111A,2020A&A...644A..75L}. The exception is S~15-19, for which we confirm an enhancement of barium, with [Ba/Fe] = 0.61. 
Based on enhancements of C and Ba and a difference of 2.4~\kms\ in radial velocity between AAS09 and their own observations, \citet{2011PASJ...63S.523H} proposed that S~15-19 is a secondary companion of the binary system where the primary companion has passed the asymptotic giant branch (AGB) evolutionary stage and carbon- and s-process-enhanced material has been transferred to the surface of S~15-19 which, thus, is a CEMP-s star. 
However, low abundance of Sr determined by \citet{2011PASJ...63S.523H} for S~15-19 ([Sr/Fe] = $-1.56$) seems to be in conflict with a hypothesis of CEMP-s star because Sr is produced by the main s-process in the AGB stars as efficiently as Ba. For example, theoretical yields from metal-poor ($Z$ = 0.0001) AGB models of 1.3, 1.5, and 2.0~$M_\odot$ predict Sr-enhancements up to [Sr/Fe] = 1 \citep[FRANEC Repository of Updated Isotopic Tables \& Yields = FRUITY database,][]{2009ApJ...696..797C,2011ApJS..197...17C}. As shown by \citet{2015ApJ...807..173H}, the [Sr/Ba] ratios in different CEMP-s stars in the Galactic halo can be of different sign, however, in every star, both Sr and Ba are enhanced relative to Fe. Another argument that further disfavors the AGB origin of Ba comes from calculations of the radial velocity ($V_{\rm rad}$) of S~15-19 observed in different years. We obtained $V_{\rm rad}$ = 225.8$\pm$1.2~\kms\ and 224.6$\pm$1.1~\kms\ for the spectra taken in 2005 and 2010, respectively. The two velocity measurements agree within the error bars.
%Radial velocity of S~15-19 is in perfect agreement with the galaxy systemic velocity 226.0$\pm$0.6~\kms\ derived by \citet{2011MNRAS.411.1013B}. 
Thus, a binarity of S15-19 is unlikely.

%the four stars, from single line in each star, from either \ or Sr\ii\ 4215~\AA. All these stars, including , are poor in Sr. 
The abundance of Sr was determined for S~04-130 and S~11-97 from Sr\ii\ 4077~\AA. For the remaining stars, their HDS/Subaru spectra are noisy in the blue spectral range, and we cannot confirm the Sr abundances, as well as the upper limits for the Eu abundance, reported by \citet{2011PASJ...63S.523H} and \citet{2020A&A...636A.111A}. 
For S~15-19, we estimated the upper limit for the yttrium abundance from Y\ii\ 5087~\AA\ (\Eexc\ = 1.08~eV, log~$gf = -0.17$), and it appears low: [Y/Fe] $< -1$. 
%, we estimated an upper limit for the yttrium abundance in S~15-19: [Y/Fe] $< -1$.
It would be extremely important to understand why Ba is strongly enhanced relative to the light n-capture elements (Sr and Y, according to HAA11 and this study, respectively) in S~15-19, given that it cannot be explained by current nuclesynthesis theory.

%  and the abundance of Zr for the two stars. We could use lines, and the Zr\ii\ 4209~\AA\ line was suitable for analysis in 

%The HDS/Subaru spectra of S~11-37, S~12-28, and S~14-98 do not cover the blue spectral range, where the  are located.  could determine abundance of Ba, 

%S15-19, compare mean from the Ba\ii\ subordinate lines ($-0.58\pm$0.15) with $-0.74$, $-0.84$, and $-1.09$ for s-process, solar mixture, and r-process. Honda+2011 adopts s-process
% there are [Sr/Ba] < 0 (down to -0.73), but all have [Sr/Fe] > and [Ba/Fe] >0; five CEMP-s stars with [Sr/Fe] = 0.07 to 1.75 and [Ba/Fe] = 0.46 to 1.80;
 
\subsection{Other elements}

In contrast to all other NLTE species, lines of Na\ione\ are strengthened in NLTE compared with LTE, resulting in negative NLTE abundance corrections. For the Na\ione\ 5895~\AA\ resonance line, $\Delta_{\rm NLTE}$ ranges between $-0.47$~dex (S~11-97: 4660/1.22/$-2.61$, [Na/Fe] = $-0.21$) and $-0.14$~dex (S~14-98: 4700/1.25/$-2.73$, [Na/Fe] = $-0.78$). Therefore, it is, in particular, important to account for the NLTE effects, when studying elemental ratios. 
For example, for S~11-97, LTE yields [Na\ione /Fe\ione] = 0.46, while a 0.62~dex lower value is obtained in NLTE.

Nickel is observed in lines of neutral atoms that are expected to be subject to overionisation like Fe\ione. Following Paper~II, we assume that the ratio of abundances derived from lines of Ni\ione\ and Fe\ione\ is nearly free of the NLTE effects and compute [Ni/Fe] as [Ni\ione /Fe\ione], using the corresponding LTE abundances.

For the two stars, S~04-130 and S~11-97, for which the UVES/VLT spectra are available, we determined the abundances of Al, Si, Zn, and Zr. The NLTE abundance of Zn was calculated using $\Delta_{\rm NLTE} = -0.11$ for Zn\ione\ 4810~\AA\ from \citet{Takeda2005zn}.

\begin{figure*}
 \begin{minipage}{170mm}
\centering
\hspace{-10mm}
	\includegraphics[width=0.25\textwidth, clip]{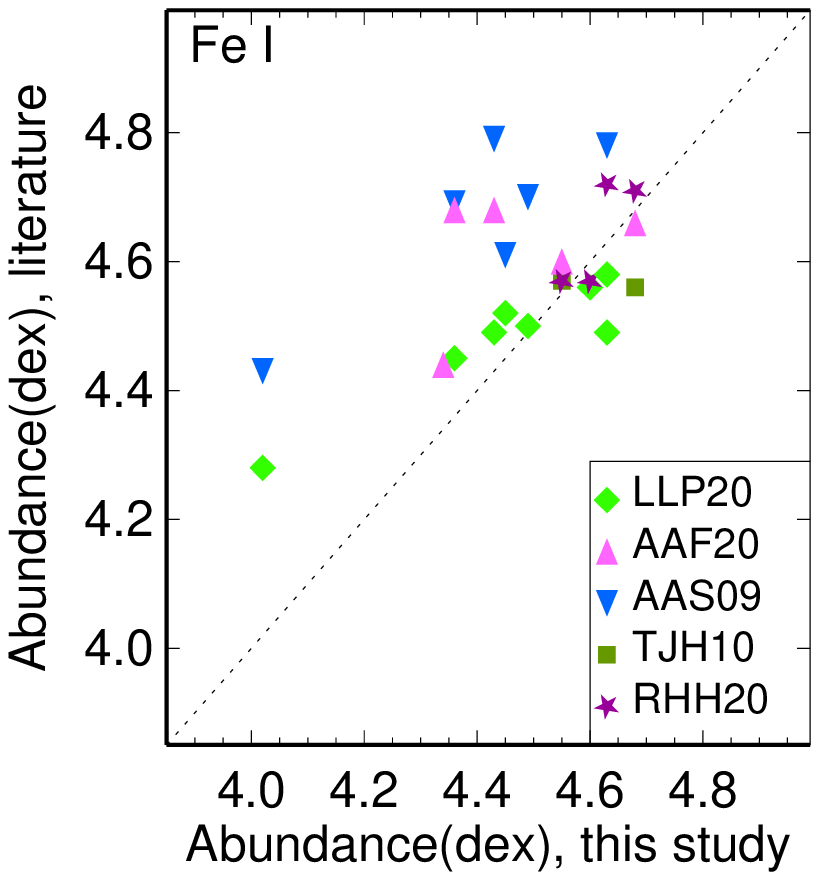}
\hspace{-5mm}
	\includegraphics[width=0.25\textwidth, clip]{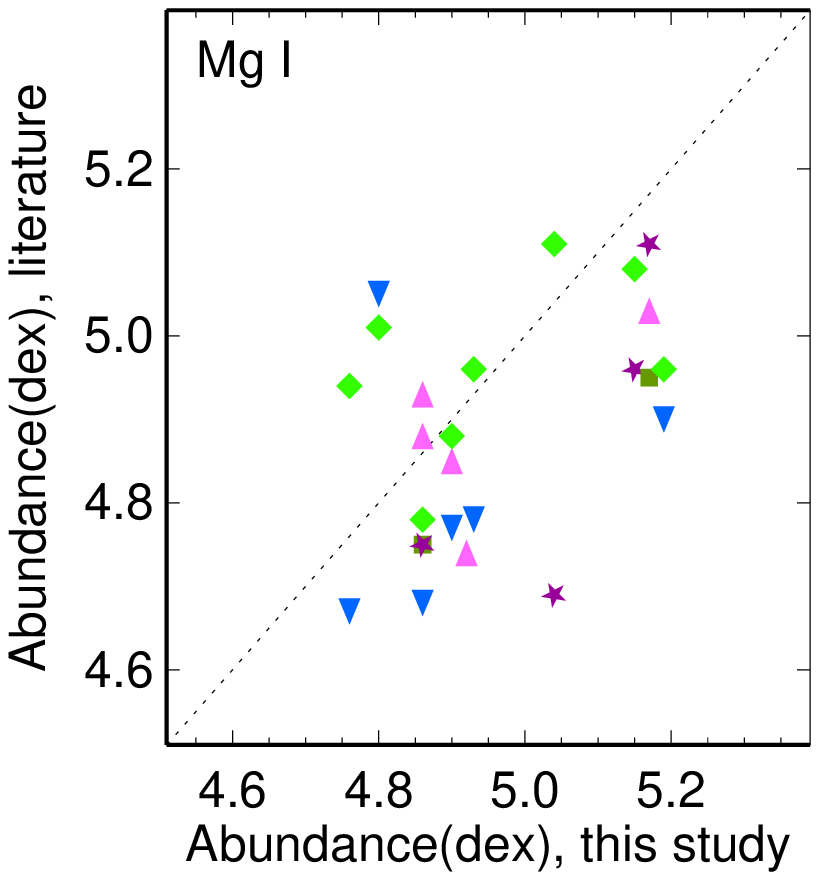}
\hspace{-5mm}
	\includegraphics[width=0.25\textwidth, clip]{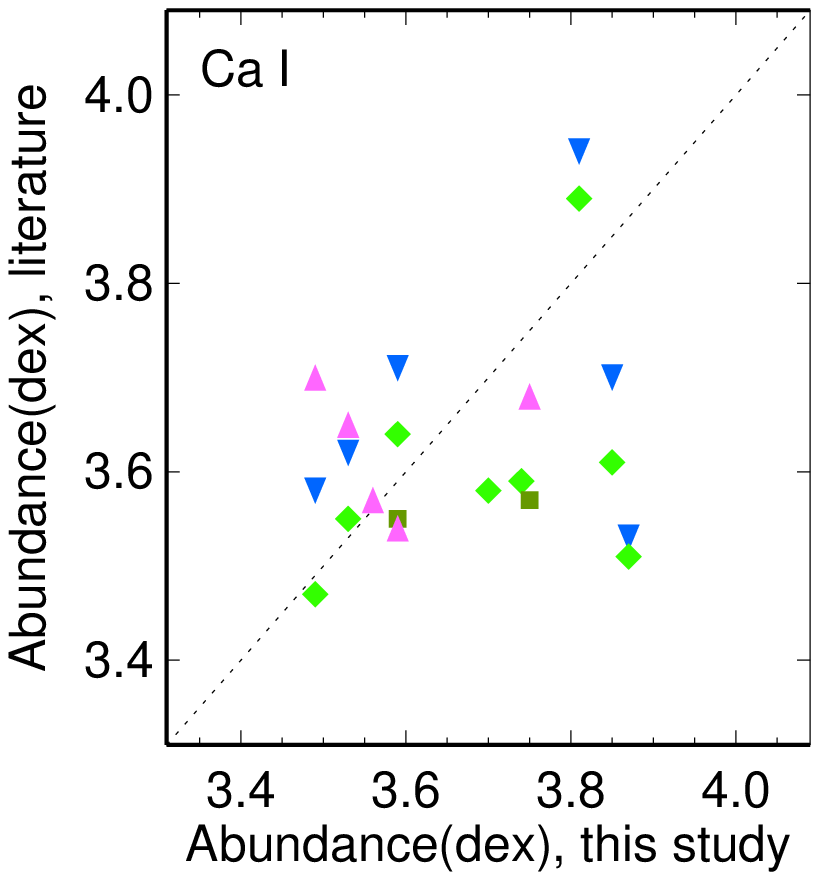}
\hspace{-5mm}
	\includegraphics[width=0.25\textwidth, clip]{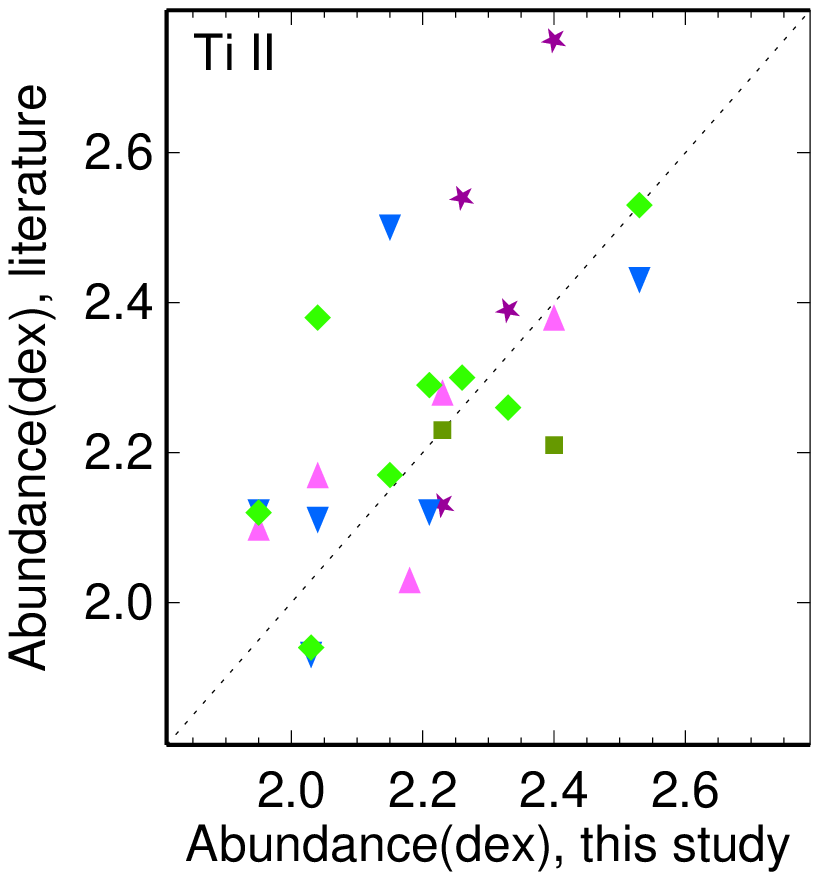}
    \caption{Mean LTE abundances ($\eps{}$) of the Sextans dSph stars derived in this study from lines of Fe\ione, Mg\ione, Ca\ione, and Ti\ii\ (panels from left to right) compared with the corresponding values from \citet[][AAS09, inverted triangles]{2009A&A...502..569A}, \citet[][TJH10, squares]{2010A&A...524A..58T}, \citet[][AAF20, triangles]{2020A&A...636A.111A}, \citet[][LLP20, rhombi]{2020A&A...644A..75L}, and \citet[][RHH20, five-pointed stars]{2020A&A...641A.127R}. }
    \label{fig:compare_abund}
\end{minipage}
\end{figure*}

\begin{figure}  %[htbp]
 \begin{minipage}{85mm}
\centering
	\includegraphics[width=0.99\textwidth, clip]{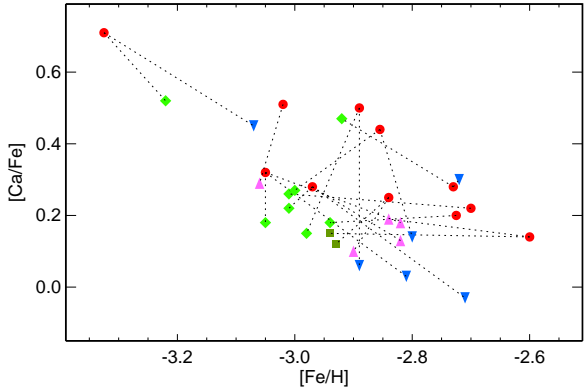}
%  \vspace{-5mm}
  \caption{\label{Fig:compare_ca} [Ca/Fe] abundance ratios for the Sextans dSph stars derived in this study (NLTE, circles) and by AAS09 (inverted triangles), TJH10 (squares), AAF20 (triangles), and LLP20 (rhombi). Common stars are connected by dotted lines.}
\end{minipage}
\end{figure}

\subsection{Influence of uncertainties in atmospheric parameters on derived elemental abundances}\label{Sect:errors}

\begin{table} % [htbp]
 \caption{\label{Tab:errors} Error budget for chemical species and elemental ratios in S~12-28.} 
 \centering
 \begin{tabular}{lcrccc}
\hline\hline \noalign{\smallskip}
Species  & $\Delta T$ & $\Delta \logg$ & $\Delta \xi$ & $\Delta$    & $\sigma$ \\
         &  $-50$ K   &   +0.1         & $-0.2$~\kms  & ($T,g,\xi$) & \\
 (1)     &     (2)    &     (3)        &  (4)         & (5)         & (6)     \\
\noalign{\smallskip} \hline \noalign{\smallskip}
 \  Na\ione &  -0.10 &             -0.01 &          0.11 & 0.15 & 0.13  \\
 \  Mg\ione &  -0.09 &             -0.03 &          0.05 & 0.11 & 0.11  \\
 \  Ca\ione &  -0.07 &             -0.02 &          0.04 & 0.08 & 0.17  \\
 \  Ti\ione &  -0.09 &             -0.02 &          0.02 & 0.09 & 0.14  \\
 \  Ti\ii   &  -0.03 &              0.02 &          0.03 & 0.05 & 0.12  \\
 \  Fe\ione &  -0.08 &             -0.02 &          0.03 & 0.09 & 0.15  \\
 \  Fe\ii   &  -0.01 &              0.02 &          0.05 & 0.05 & 0.10   \\
 \  Ni\ione &  -0.08 &             -0.01 &          0.04 & 0.09 & 0.1  \\
 \  Ba\ii   &  -0.04 &              0.01 &          0.07 & 0.08 & 0.1  \\
 \ [Fe/H]   &  -0.04 &              0.00 &          0.03 & 0.05 & 0.18 \\
 \ [Na/Fe]  &  -0.06 &             -0.01 &          0.08 & 0.10 & 0.22 \\
 \ [Mg/Fe]  &  -0.05 &             -0.03 &          0.02 & 0.06 & 0.21 \\
 \ [Ca/Fe]  &  -0.03 &             -0.02 &          0.01 & 0.04 & 0.25 \\
 \ [TiII/Fe] &  0.01 &              0.02 &          0.00 & 0.02 & 0.22 \\
 \ [NiI/FeI] &  0.00 &              0.01 &          0.01 & 0.01 & 0.18 \\
 \ [Ba/Mg]  &   0.05 &              0.04 &          0.02 & 0.07 & 0.15 \\
\noalign{\smallskip} \hline
\end{tabular}
\end{table}

 We examined the uncertainties linked to our choice of atmospheric parameters. Since our sample stars have close together effective temperatures, surface gravities, and microturbulent velocities, we selected one of them, S~12-28 (4590/0.95/$-2.86$, $\xi_t$ = 2.1~\kms), for calculations.
The uncertainties in derived elemental abundances and abundance ratios were evaluated by varying $\Teff$ by $-50$~K, $\logg$ by 0.1~dex, and $\xi_t$ by $-0.2$\kms\ in the stellar atmosphere model. 
Table~\ref{Tab:errors} summarizes the various sources of uncertainties. The quantity $\Delta (T,g,\xi)$ listed in Col.~5 is the total impact of varying each of the three parameters, computed as the quadratic sum of Cols. 2-4. For comparison, Table~\ref{Tab:errors} indicates also the stochastic errors $\sigma$. For the elemental ratios, which are the main object of the galactic chemical evolution study in Sect.~\ref{sect:trends}, $\Delta (T,g,\xi)$ is substantially smaller than the corresponding $\sigma$. 

\subsection{Comparisons with the previous studies}\label{Sect:others}

In order to check an influence of differences in the measured EWs, the derived atmospheric parameters, and the used LTE codes on the computed elemental abundances, 
we compared the absolute LTE abundances obtained in this study from lines of Fe\ione, Mg\ione, Ca\ione, and Ti\ii\ with the previous measurements for common stars (Fig.~\ref{fig:compare_abund}). 
The paper by \citet{2020AA...642A.176T} was not included in these abundance comparisons because we have the only star in common, S~11-97, for which \citet{2020AA...642A.176T} do not provide abundances of Mg and Ti and the Ca abundance is based on only one line. 
An impression of systematically higher Fe\ione -based abundances compared with our values is mostly produced by the data of AAS09, while the abundance differences with the remaining studies do not exceed 0.1~dex for most of the stars. Discrepancies with AAS09 for Fe\ione, by up to 0.4~dex, cannot be explained by the differences in EWs or $\Teff$. Furthermore, the Mg\ione -based abundances of AAS09 are lower compared with our values, by 0.15~dex, on average. For each chemical species, the data reveal a scatter at the level of 0.3~dex. There are also outliers. 
For example, for the star S~04-130, with the largest difference in $\Teff$ between this study and RHH20, we obtained higher abundance from lines of Mg\ione, by 0.35~dex, but surprisingly small discrepancy for the Fe\ione -based abundance. 
%in  some cases, large abundance differences seen in  Fig.~\ref{fig:compare_abund} cannot be explained by the discrepancies in EWs (Fig.~\ref{Fig:subaru}) and atmospheric parameters (Fig.~\ref{Fig:comp1}).
% For example, for S~15-19, we obtained lower Fe\ione -based abundance compared with that in HAA11 and LLP20, by 0.41~dex and 0.26~dex, respectively. For S~11-37, our Ca abundance is higher than that in AAS09 and LLP20, by 0.34~dex and 0.36~dex, respectively. Our Ti\ii -based abundances are lower than those in AAS09 for S~11-37, by 0.35~dex, and in RHH20 for S~04-130, by 0.27~dex. 

Next, we checked the differences in the [Ca/Fe] versus [Fe/H] trends and how they affect conclusions on chemical enrichment of Sextans. Calcium is the best observed $\alpha$-element. Furthermore, [Ca/Fe] is the least affected by the departures from LTE and differences in $\Teff$ because lines of Ca\ione\ and Fe\ione\ have similar NLTE abundance corrections and similar sensitivity to temperature variations. For each star, Fig.~\ref{Fig:compare_ca} displays all the [Ca/Fe] values available in the literature (all derived under the LTE assumption) together with our NLTE abundance ratio. 

Surprisingly, for five of six stars, the LTE values of [Fe/H] in AAS09 are higher than our NLTE abundances. As a result, AAS09 obtained the narrower metallicity range, $-2.9 <$ [Fe/H] $< -2.7$, and the lower [Ca/Fe] values, which are close to the solar one. 

Our stellar sample largely overlaps wih that of LLP20, however, we determined essentially broader metallicity range covered ($-3.05 \le$ [Fe/H] $\le -2.61$ except the most MP star at [Fe/H] = $-3.32$) compared with that of LLP20 ($-3.05 <$ [Fe/H] $< -2.9$). The discrepancies in [Fe/H] can be understood. First, the NLTE abundances from lines of Fe\ione\ are higher than the LTE abundances, by up to 0.2~dex. Second, we relied on the photometric effective temperatures that are higher compared with the spectroscopic ones in LLP20. This also leads to higher abundances from lines of Fe\ione. We remind that, in  this study, the NLTE abundances from lines of Fe\ione\ and Fe\ii\ are consistent in each star. The differences in [Ca/Fe] are smaller than that for [Fe/H] and do not exceed 0.2~dex for seven of eight common stars.

%Unexpectedly, for the three of six stars, their LTE metallicities in AAS09 are substantially higher than our NLTE abundances. This explains the lower [Ca/Fe] values of AAS09 compared with our results.
 
For every star, LLP20 obtain lower [Fe/H] compared with that in AAS09 and, as result, higher [Ca/Fe]. The difference in [Fe/H] ranges between $-0.1$ and $-0.4$~dex and amounts to, on average,  $-0.2$~dex. In the view of moderate discrepancies in $\Teff$ ($-65$~K) between LLP20 and AAS09 and using in both studies the LTE abundances from the Fe\ione\ lines, it is difficult to explain the large differences in the derived [Fe/H]. 

\begin{figure}  %[htbp]
 \begin{minipage}{90mm}
\centering
	\includegraphics[width=0.99\textwidth, clip]{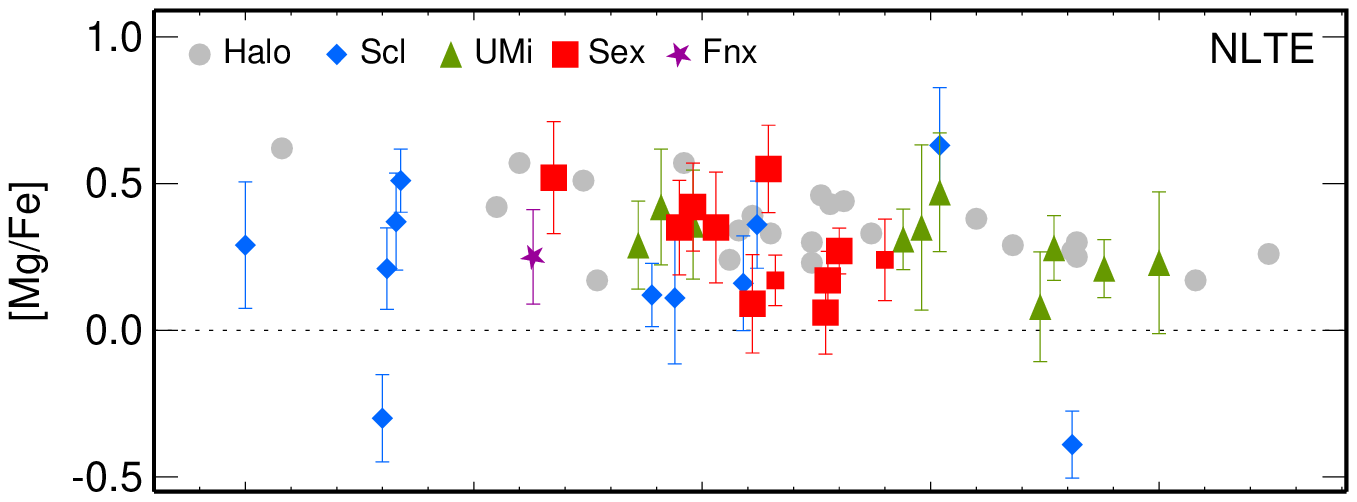}

  \vspace{-12mm}
	\includegraphics[width=0.99\textwidth, clip]{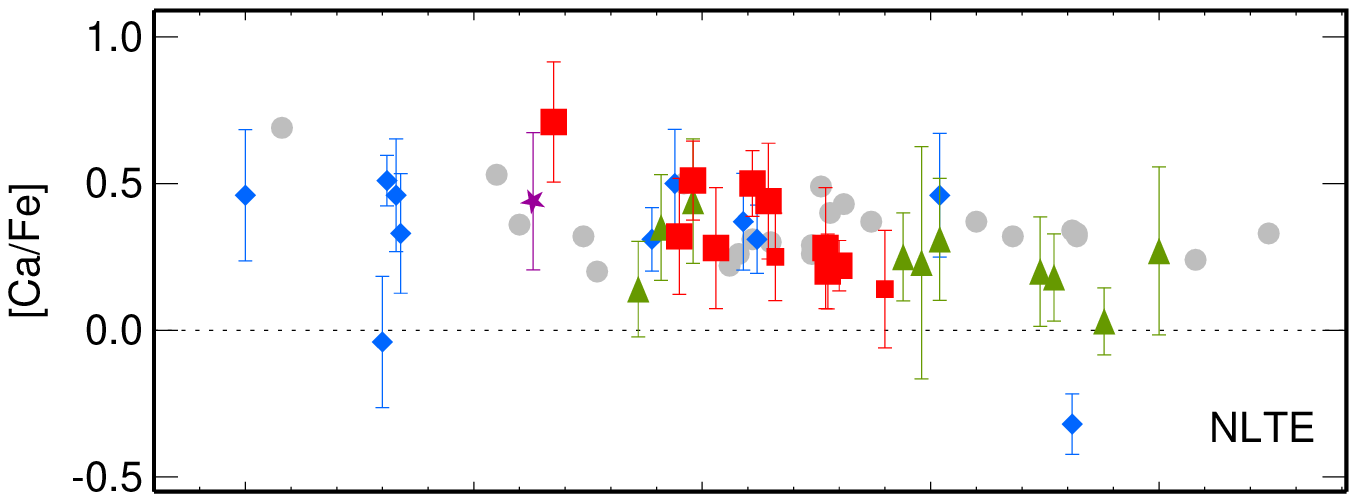}

  \vspace{-12mm}
	\includegraphics[width=0.99\textwidth, clip]{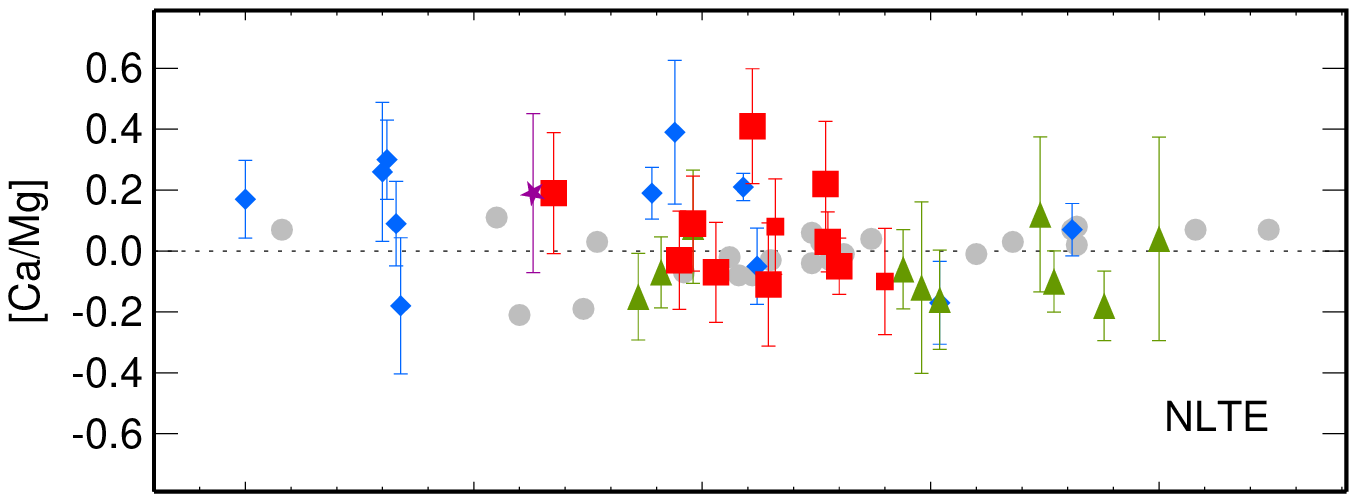}

  \vspace{-12mm}
	\includegraphics[width=0.99\textwidth, clip]{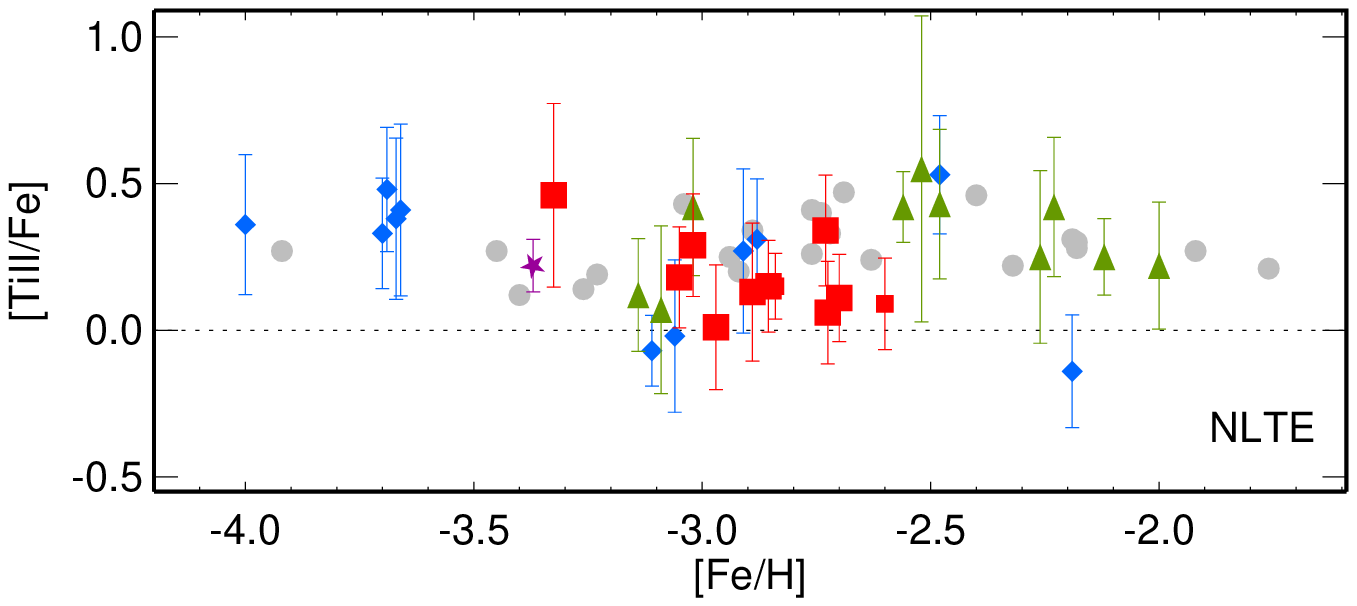}
% \hspace{-6mm}  
%  \vspace{-13mm}
%  \vspace{-5mm}
%  \resizebox{88mm}{!}{\includegraphics{mgca_nlte_final.ps}}
  \caption{\label{Fig:alpha} Stellar elemental ratios from the NLTE calculations for the $\alpha$-process elements Mg, Ca, and Ti in the Sextans dSph (squares) compared with that derived in Paper~II for the classical dSphs Sculptor (rhombi), Ursa Minor (triangles), and Fornax (five-pointed star) and the MW halo (circles). The smaller size squares correspond to the Sextans stars S~11-04 and S~24-72 analysed in Paper~II. 
%The error bars were computed as $\sigma_{\rm X-Y} = \sqrt{\sigma_{\rm X}^2 + \sigma_{\rm Y}^2}$. 
}
\end{minipage}
\end{figure}

\section{Abundance trends in Sextans}\label{sect:trends}

\begin{figure}  %[htbp]
 \begin{minipage}{90mm}
\centering
	\includegraphics[width=0.99\textwidth, clip]{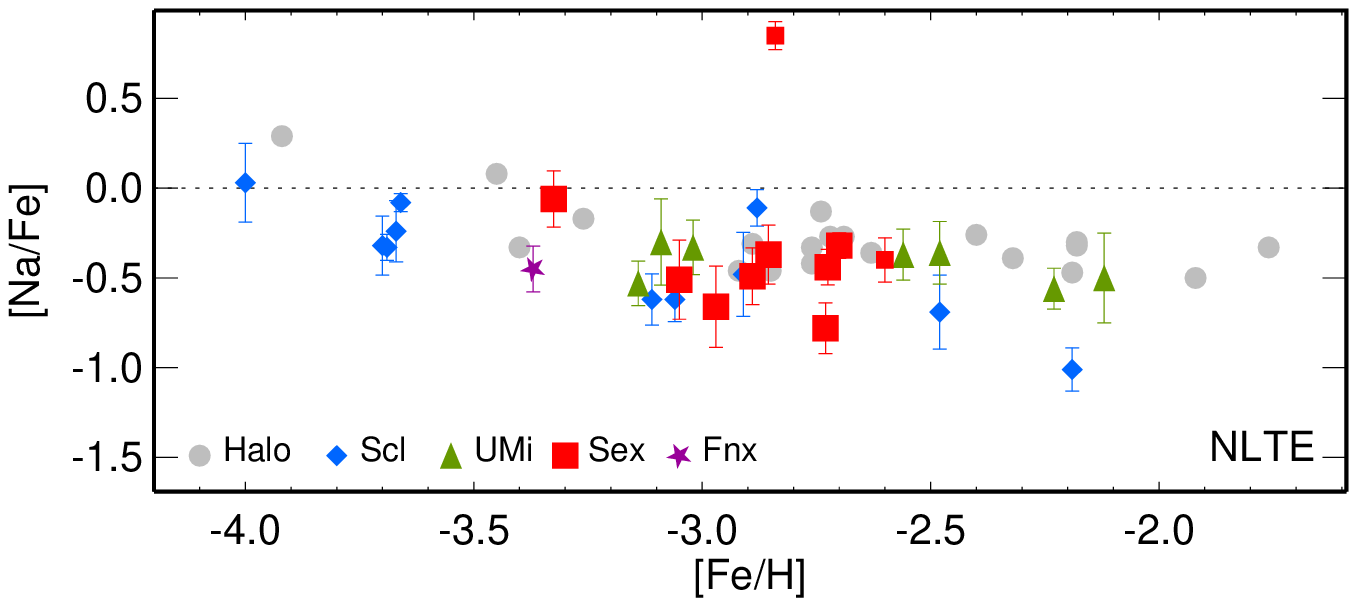}

  \vspace{-5mm}
	\includegraphics[width=0.99\textwidth, clip]{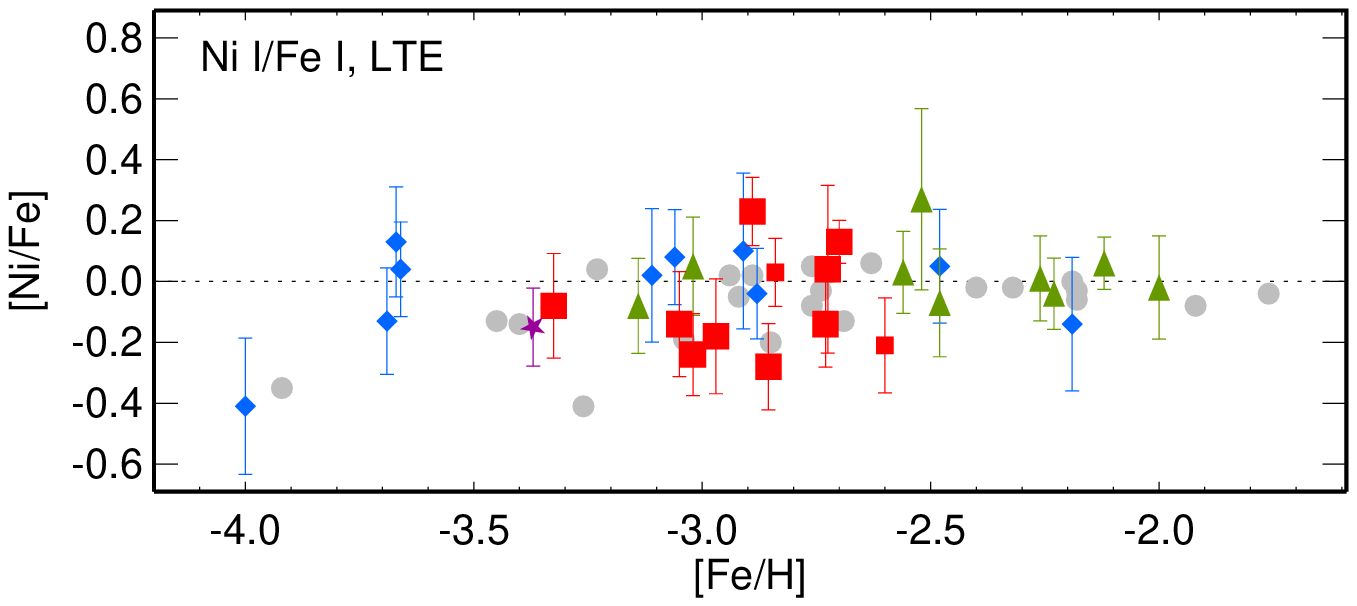}
%  \vspace{-5mm}
  \caption{\label{Fig:nafe} The same as in Fig.~\ref{Fig:alpha} for [Na/Fe] (top panel) and [Ni/Fe] (bottom panel). }
\end{minipage}
\end{figure}

We inspect the abundance trends in the $-3.3 <$ [Fe/H] $< -2.6$ metallicity range of the Sextans dSph  using eleven stars with the abundances measured for, at least, seven chemical elements in each star.
Using accurate atmospheric parameters, accounting for the departures from LTE for the elemental ratios, which involve the chemical species with different and metallicity dependent magnitudes of the NLTE effects, 
%in the stellar sample, covering a broad metallicity range, in particular, VMP regime, 
increases a credit of confidence in the inferred picture of chemical enrichment of the Sextans dSph.%We present the NLTE abundance ratios for the largest of ever analysed samples of the VMP stars in the Sextans dSph and focus on the elemental ratios obtained for more than half sample stars. 
The abundance trends determined for Sextans are compared with the corresponding trends in the other two classical dSphs, Sculptor and Ursa Minor, and in the MW halo, as obtained in Paper~II using the same methods of abundance determinations. The error bars in Figs.~\ref{Fig:alpha}-\ref{Fig:bamg} were computed with taking into account the stochastic abundance errors of the species involved in a given elemental ratio (that is the abundance difference in logarithmic scale, [X/Y] = [X/H] -- [Y/H]): $\sigma_{\rm X-Y} = \sqrt{\sigma_{\rm X}^2 + \sigma_{\rm Y}^2}$. As discussed in Sect~\ref{Sect:errors}, the abundance uncertainties caused by the uncertainties in atmospheric parameters are substantially smaller than the stochastic errors. 

\subsection{ $\alpha$-process elements}

\begin{table} % [htbp]
 \centering
 \caption{\label{Tab:mgca} [Mg/Fe] and [Ca/Fe] mean NLTE abundance ratios for different metallicity ranges of the Sextans dSph and the MW.} 
 \begin{tabular}{lccc}
\hline\hline \noalign{\smallskip}
  \multicolumn{1}{r}{[Fe/H] =}  &      $-3.2$ to $-2.6$ & $-3.2$ to $-2.8$ &  $-2.8$ to $-2.6$ \\
\hline \noalign{\smallskip}
         & \multicolumn{3}{l}{[Mg/Fe]} \\
 Sextans & 0.27 $\pm$ 0.08 (10) &   0.32 $\pm$ 0.09 (6) &   0.18 $\pm$ 0.08 (4) \\
 MW      & 0.37 $\pm$ 0.07 (11) &   0.37 $\pm$ 0.08 (5) &   0.36 $\pm$ 0.07 (6) \\
         & \multicolumn{3}{l}{[Ca/Fe]} \\
 Sextans & 0.31 $\pm$ 0.06 (10) &   0.38 $\pm$ 0.06 (6) &   0.21 $\pm$ 0.05 (4) \\
 MW      & 0.35 $\pm$ 0.05 (11) &   0.32 $\pm$ 0.06 (5) &   0.37 $\pm$ 0.06 (6) \\
\noalign{\smallskip}\hline \noalign{\smallskip}
\multicolumn{4}{l}{{\bf Notes.} The errors are computed as $\sqrt{\sigma_{\rm mean}^2 + \Delta(T,g,\xi)^2}$. } \\
\multicolumn{4}{l}{The numbers in parentheses are the numbers of stars.} \\
\noalign{\smallskip} \hline
\end{tabular}
\end{table}

%{\it $\alpha$-process elements.}
The stars in Sextans reveal similar enhancements of Mg and Ca relative to iron (Fig.~\ref{Fig:alpha}), and 
%As shown, each star reveals enhancements of Mg and Ca relative to iron, like that for the VMP regime of the Sculptor and Ursa Minor dSphs and the MW halo. However, 
we find a hint of a decline in $\alpha$-enhancement for [Fe/H] $> -2.8$, where both [Mg/Fe] and [Ca/Fe] decrease from about 0.4~dex to 0.2~dex. 
The [Ti/Fe] versus [Fe/H] diagram is less informative because of bigger uncertainty in the data compared with that for [Mg/Fe] and [Ca/Fe]. 
To be precise, we computed the mean elemental ratios for the stars lying below and above [Fe/H] $= -2.8$ (Table~\ref{Tab:mgca}). The most MP star at [Fe/H] $= -3.32$ was not included in calculations of the mean, although its including would strengthen our conclusions. For comparison, Table~\ref{Tab:mgca} presents the mean elemental ratios for the same metallicity ranges in the MW, using the NLTE abundances from Paper~II. In Sextans, the differences in [Mg/Fe] and [Ca/Fe] between the [Fe/H] $< -2.8$ and $> -2.8$ stars amount to 0.14~dex and 0.17~dex, respectively. They exceed the total uncertainty produced by the stochastic ($\sigma_{\rm mean}$) and systematic ($\Delta(T,g,\xi)$) errors, by a factor of 1.5 and 3 for [Mg/Fe] and [Ca/Fe], respectively. Here, $\sigma_{\rm mean}$ is the standard error of the mean. In the $-3.2 <$ [Fe/H] $< -2.8$ metallicity range, Sextans is enhanced in Mg and Ca at the same level as our Galaxy. This suggests an early enrichment of this small galaxy by massive stars in numbers following classical initial mass function (IMF). For the less MP ($-2.8 <$ [Fe/H] $< -2.6$) stars, [Mg/Fe] and [Ca/Fe] in Sextans are lower compared with the corresponding numbers in the MW, by 0.18~dex and 0.16~dex, which exceed the uncertainty in the derived elemental ratios, by a factor of more than two.

Usually, a decline of [$\alpha$/Fe] is explained by increased production of Fe in the SNe~Ia appeared. From the analysis of the colour-magnitude diagram \citet{2009ApJ...703..692L} find a timescale of about 4~Gyr for star formation in the central region of Sextans. This is enough time for SNe~Ia to appear. However, the question is how much iron was produced by massive stars before the first explosions of SNe~Ia. To our knowledge, none of the dSphs studied in
the literature has the onset of iron production by SNe~Ia at metallicity below [Fe/H] = $-2.5$. Could a decrease in [$\alpha$/Fe] be explained differently? 
The sampling of the IMF in small systems such as dwarf galaxies can be challenging as seen in high-resolution chemo-dynamical simulations \citep[for example,][]{2018A&A...616A..96R}. There is always a possibility that some massive core-collapse stars are missing at some stage of their chemical evolution or in their building blocks. This scenario was also put forward for low [$\alpha$/Fe] stars in the Sculptor dSph \citep[for example,][]{2015A&A...583A..67J}.
%High-resolution chemo-dynamical simulations of dwarf galaxies in a $\Lambda$CDM cosmology predict that dark haloes containing stars and having different IMF can be accreted by the galaxy . The accreted stars may have different [$\alpha$/Fe] compared with the stars of close metallicity in the accreting galaxy.}

 Another explanation of the early knee on the [$\alpha$/Fe] trends would be an accretion/merger event. Anomalies in the spatial distribution and kinematic properties of Sextans stellar population and clear signs of accretion experienced by the galaxy are uncovered by \citet{2018MNRAS.480..251C}. They expanded the study of \citet{2011MNRAS.411.1013B} who detected that Sextans' central regions at projected radii $R < 0.^\circ22$ are more metal rich and have smaller velocity dispersion than the outer parts. Cosmological simulation of \citet{2016MNRAS.456.1185B} predicts a scenario where a dSph forms `outside-in': 'an old, metal-poor population results from a first episode of star formation', while 'the younger component forms when a late accretion event adds gas and reignites star formation'. The younger population is concentrated at the centre of galaxy, while the old population is dispersed by the mergers. \citet{2020A&A...641A.127R} use observational findings of \citet{2018MNRAS.480..251C} and ideas of \citet{2016MNRAS.456.1185B} to explain existing the two knees in the Sextans dSph, as found from their analysis of [Mg/Fe] in 41 stars, at [Fe/H]$_{\rm knee} = -2$ and [Fe/H]$_{\rm knee} = -2.5$. The latter is consistent with our finding, when taking into account the uncertainty of 0.3~dex in metallicity obtained by \citet{2020A&A...641A.127R}.
%As expected from comparisons in Sect.~\ref{Sect:others}, our $\alpha$-abundance trends are different from those reported by AAS09: we did not find a close-to-solar $\alpha$/Fe abundance ratio for [Fe/H] $< -2.6$. Our results support the LLP20 study only in part, namely, the Sextans [Fe/H] $ < -2.8$ population follows the MW halo-like plateau at [$\alpha$/Fe] $\sim$ 0.4. Different $\alpha$-abundance trend was obtained in this study for less metal-poor stars. 
The knee at [Fe/H] $\simeq -2.0$ can be connected with the younger population, whereas the more metal-poor knee may be formed before the accretion. This does not contradict the results of \citet{2020AA...642A.176T} who find the knee at [Fe/H]$_{\rm knee} \simeq -2$ for the stellar sample in the central region of Sextans, at projected radii $R < 0.^\circ2$ (see their Fig.~1).

\subsection{Sodium and nikel }

Abundance trends for Na and Ni are similar to that for the other galaxies. 
%{\it Sodium.} 
As shown in Paper~II, accounting for the NLTE effects leads to fairly similar [Na/Fe] trends in different mass galaxies. The Sextans dSph is not exception. We obtained that Na is underabundant relative to Fe for [Fe/H] $> -3.2$, at the level of 0.5~dex (Fig.~\ref{Fig:nafe}). The exception is a carbon-enhanced star S~24-72, which is strongly enhanced also in Na. However, our second carbon-enhanced star, S~15-19, is not exceptional, and its close-to-solar [Na/Fe] lies well on the rising trend of [Na/Fe] with decreasing metallicity observed in the MW halo and Sculptor dSph for [Fe/H] $< -3.2$. 

Nikel follows Fe (Fig.~\ref{Fig:nafe}), although with the larger scatter of [Ni/Fe] compared with that for the other galaxies in the overlapping metallicity range. The mean [Ni/Fe] = $-0.08\pm0.17$.
%[Ni/FeI] = 0.06 and 0.12 for S04-130 and S~11-97; -0.27 and 0.08 for S~11-04 and S24-72. Mean [Ni/FeI] = -0.13+-0.09 for 7 Subaru stars. 

\begin{figure}  %[htbp]
 \begin{minipage}{90mm}
\centering
	\includegraphics[width=0.99\textwidth, clip]{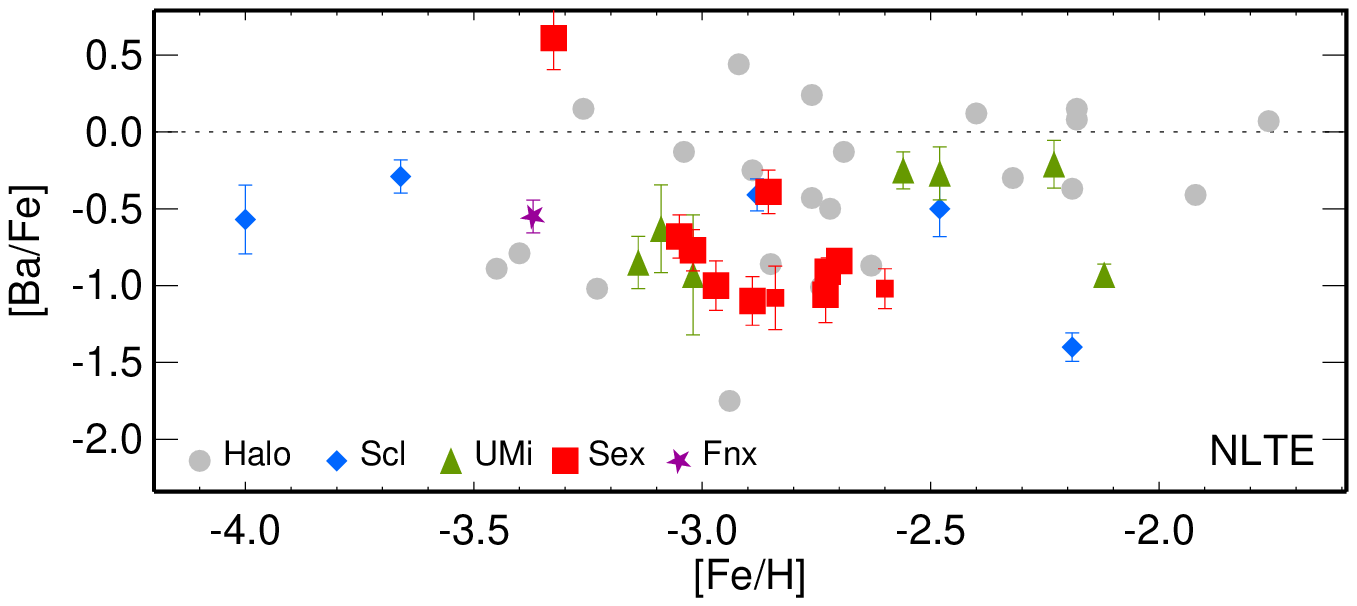}

  \vspace{-5mm}
	\includegraphics[width=0.99\textwidth, clip]{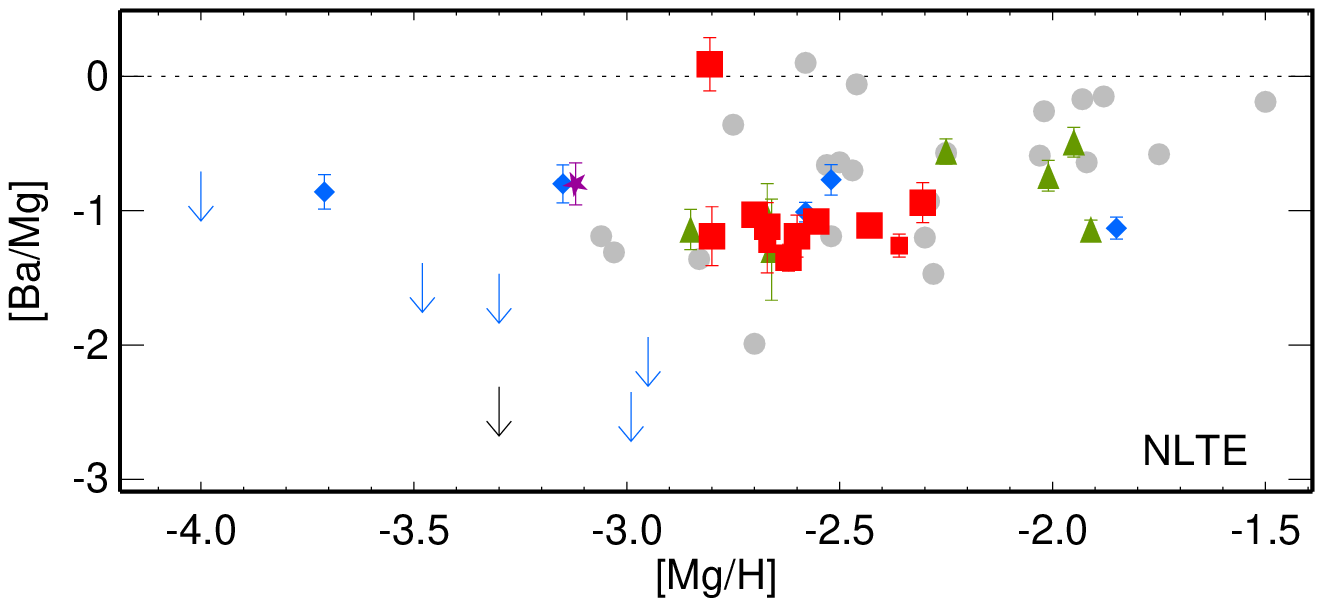}
%  \vspace{-5mm}
%	\includegraphics[width=0.99\textwidth, clip]{srba_nlte_sex_class.ps}
%  \vspace{-5mm}
  \caption{\label{Fig:bamg} The same as in Fig.~\ref{Fig:alpha} for [Ba/Fe] versus [Fe/H] (top panel) and [Ba/Mg] versus [Mg/H] (bottom panel). }
\end{minipage}
\end{figure}

%Ba/H   Sr/Ba : Sr/Fe
%%-3.64  0.52  -0.47
%-3.55  0.14   -0.80
%-3.79  0.01   -0.76 - no Sr
%-3.92  0.65   -0.43
%-3.62  0.51   -0.51
%-2.71  -1.67  -1.06 - no Sr

\subsection{Neutron-capture elements}
%{\it Neutron-capture elements.} 

Similarly to the other dSphs, Sextans is poor in Ba in the VMP regime (Fig.~\ref{Fig:bamg}). The derived [Ba/Fe] ratios are close to the [Ba/Fe] floor of the MW halo. The exception is a carbon-enhanced star S~15-19. Since an enhancement in Ba is not accompanied by an enhancement in Sr, S~15-19 cannot be a CEMP-s star (see Sect.~\ref{sect:heavy}). Such a puzzle of S~15-19, that is high Ba, but low Sr abundance, needs further investigation.

As noticed in Paper~II, for better understanding the evolution of Ba and Sr in galaxies, it is useful to consider their abundances relatively to Mg. This removes any potential pollution of the iron abundances by the ejecta of SNe~Ia.
%that can be exactly the case in the Sextans dSph, even for very low metallicities. 
Except S~15-19, the remaining Sextans stars reveal an extremely tight relation between Ba and Mg (Fig.~\ref{Fig:bamg}), with the mean [Ba/Mg] = $-1.19$ ($\sigma_{\rm mean}$ = 0.03~dex). This suggests a common origin of Ba and Mg in massive stars and excludes the main s-process associated with the thermally pulsing asymptotic giant branch phase of intermediate-mass stars (2--4 $M_\odot$) as the source of Ba. The low-metallicity and fast rotating massive stars can produce Ba in the s-process, as proposed, for example, by \citet{2008ApJ...687L..95P} and \citet{2014A&A...565A..51C}, however, the simulations predict a substantial scatter of [Ba/Fe] that is not observed in Sextans (top panel in Fig.~\ref{Fig:bamg}). There seems to be a contradiction with analysis of the Ba\ii\ resonance and the subordinate lines that favour the s-process origin of Ba in the Sextans dSph. 
 An alternate is the r-process. Its astrophysical sites are still debated \citep[see][for a thorough review]{2021RvMP...93a5002C}. In order to produce the flat [Ba/Mg] ratio, the Ba production site should act on the timescales of standard SNe~II. \citet{2021RvMP...93a5002C} estimate a frequency of neutron-star and neutron-star / black hole (binary) mergers as well as magneto-rotational supernovae with jets (MHD-SNe) as 1 event per 100 to 1000 regular SNe~II. 
%Therefore, such rare events are unlikely to happen in the small galaxy like the Sextans dSph. 
Further galactic chemical evolution models can shed light into the r-process site in Sextans, when the galaxy was very metal-poor. 
 In the overlapping Mg abundance range, the other two classical dSphs, in Sculptor and Ursa Minor, have very similar Ba/Mg abundance ratio, suggesting a common astrophysical site for Ba production. This ratio can serve to constrain models of Ba synthesis in the r-process. 
%This seems to be in conflict with a hint of the s-process origin of Ba in the Sextans dSph that was discussed in Sect.~\ref{sect:heavy}. However, there is no conflict, if the s-nuclei can be produced efficiently in the low-metallicity and fast rotating massive stars \citep[see, for example,][]{2008ApJ...687L..95P,2019MNRAS.489.5244R}. 

We have reliable measurements of the Sr abundance for the four stars in the Sextans dSph. Each of them is poor in Sr. At the same time, Sr is enhanced relative to Ba (Table~\ref{Tab:abund}), such that the three of four stars lie on a downward trend of [Sr/Ba] with [Ba/H] reported earlier for the MW halo \citep[for example,][]{Honda2004,Francois2007} and the classical dSphs \citep{2017A&A...608A..89M}. 
%The two stars have close-to-solar [Sr/Ba]. 

\section{Conclusions}\label{sect:Conclusions}

In this paper, we extend a homogeneous set of accurate NLTE abundances for the Galaxy dwarf satellites that was established in our earlier studies \citep{dsph_parameters,2017A&A...608A..89M,2019AstL...45..259P,2021_coma} with the purpose of providing the community with a robust
framework for galactic chemical evolution studies.
We assembled a sample of eleven VMP stars belonging to the Sextans dSph, for which high-resolution spectra are available. For two of them, S~11-04 and S~24-72, the NLTE abundance analysis was performed in Papers~I and II. 
%This is largest of ever analysed samples of the VMP stars in this galaxy. 

In order to obtain as accurate elemental abundances as possible, we have taken the following steps.

{\it High-resolution spectroscopy.} For seven stars, the raw data taken from the HDS/Subaru archive have been re-reduced. For the remaining two stars, we used the reduced UVES/VLT spectra.

{\it Atmospheric parameters.} The effective temperatures and surface gravities were derived from the spectral energy distributions and the known distance of the Sextans dSph. Metallicities (Fe abundances) are based on the NLTE calculations for lines of Fe\ione\ and Fe\ii. The semi-empirical formula from in Paper~I was applied to obtain the microturbulent velocities.

{\it The NLTE line formation} calculations were performed to determine abundances of Na, Mg, Ca, Ti, Fe, and Ba for each star. Abundance of Ni was evaluated under the LTE assumption. For the two stars with the UVES/VLT spectra available, we also determined the LTE and NLTE abundances of Al, Si, Zn, Sr, and Zr.

The obtained results can be summarised as follows.

\begin{itemize}
\item Our NLTE analysis finds a
%We determined essentially 
broader metallicity range, $-3.32 \le$ [Fe/H] $\le -2.61$, compared to the previous LTE studies with the same stellar sample \citep{2009A&A...502..569A,2020A&A...644A..75L}.
\item The sample stars reveal similar enhancements of Mg and Ca relative to iron. We found the differences in $\alpha$-enhancement between the [Fe/H] $< -2.8$ and $> -2.8$ stars, at the level of 2-3$\sigma_{\rm mean}$, which is meaningfull. The lower metallicity stars in the Sextans dSph reveal the plateau at [Mg/Fe] = 0.32 and [Ca/Fe] = 0.38, which are very similar to that for the MW halo. The [Fe/H] $> -2.8$ stars in Sextans have lower [Mg/Fe] and [Ca/Fe], by 0.14~dex and 0.17~dex, respectively. This is a hint of the knee at [Fe/H] $\sim -2.8$. 
The obtained results provide a firm evidence for that an early enrichment of this small galaxy in the $\alpha$-elements and Fe was produced by massive stars in numbers following classical initial mass function and provide a hint of the onset of Fe production by SNe~Ia at very low metallicity, [Fe/H] $\sim -2.8$. None of the classical dSphs studied so far has shown a knee below [Fe/H] = $-2.5$. Complex behavior of $\alpha$/Fe in Sextans further demonstrates the difference between the evolution of dwarf galaxies and the MW. 

\item The Sextans galaxy is poor in Ba in the VMP regime. The new result is an extremely tight relation  between Ba and Mg, suggesting their common origin in massive stars and production of Ba on the timescales of standard supernovae, probably in the r-process events.
\item We confirm that a carbon-enhanced star S~15-19 is strongly enhanced also in Ba, with [Ba/Fe] = 0.61. This star is unlikely to be a CEMP-s star because of non-detection of variation in the radial velocity and very low abundances of the light n-capture elements Sr and Y that are produced in the s-process as efficiently as Ba. Such a n-capture element puzzle deserves further detailed investigation from both observational, and theoretical side. 
%It would be extremely important to understand a strong enhancement of the heavy n-capture element Ba relative to the light n-capture elements Sr and Y.
\item The Sextans stars in the VMP regime are deficient in Na, except a carbon-enhanced star S~24-72 that is strongly enhanced also in Na, and form a trend of [Na/Fe] with metallicity that is indistinguishable from that in the Milky Way halo and the Sculptor and Ursa Minor dSphs. This suggests that the process of Na synthesis (carbon burning process) was identical in all systems, independent of their mass.
\item The Ni/Fe ratios scatter around the solar value, similarly to that for the Milky Way halo and a sample of six dSphs, as obtained earlier in Paper~II.
\end{itemize}

{\it Acknowledgments.} 
This study is based in part on data collected at Subaru Telescope and obtained from the SMOKA, which is operated by the Astronomy Data Center, National Astronomical Observatory of Japan.
This research has made use of the VALD and ADS\footnote{http://adsabs.harvard.edu/abstract\_service.html} databases and the VizieR catalogue access tool, CDS, Strasbourg, France (DOI: 10.26093/cds/vizier). The original description of the VizieR service was published in \citet{2000A&AS..143...23O}. S.A.Y. and A.K.B. gratefully acknowledge support from the Ministry of Education of the Russian Federation (project FSZN-2020-0026). 

\section{Data availability}

The data underlying this article will be shared on reasonable request to the corresponding author.

\bibliography{atomic_data,mashonkina,nlte,method,mp_stars,scl,references,sr2020,vizier.bib}
%\bibliography{scl}
\bibliographystyle{mnras}

\label{lastpage}
\end{document}